%% file: petefest6.tex
\input{epsf}
\documentstyle[prd,eqsecnum,aps,twocolumn]{revtex}
\input{title2.tex}

\input{psfig}
\begin{document}
\draft
\preprint{LA-UR-96-4514} 
\title{From Landau's Hydrodynamical Model  \\to Field Theory
Models of Multiparticle Production: \\
 A Tribute to Peter Carruthers on his 61st Birthday}
\author{
Fred Cooper \\
{\small \sl Theoretical Division, Los Alamos National Laboratory,}\\  
{\small \sl Los Alamos, NM 87545}\\
}
\date{\today}
\def\ddk{[d^3{\rm k}]}
\def\kk{{\rm k}}
\def\baselinestretch{1.0}
\def\ddk{[d^3{\rm k}]}
\def\kk{{\rm k}}

\abstract{
We review the assumptions and domain of applicability of Landau's
Hydrodynamical Model. By considering two models of particle production,
pair production from strong  electric fields and particle
production in the linear $\sigma$ model, we demonstrate that many of
Landau's ideas are verified in explicit field theory calculations.
}
\maketitle2 
\narrowtext

\section{Introduction}

In the early 1970's Peter Carruthers \cite{pete1} \cite{pete2} (at the
prodding of one of his graduate students Minh Duong Van) realized that
Landau's Hydrodynamical model \cite{landau1} explained the single
particle rapidity distribution $dN/d \eta$ of pions produced in the
inclusive reaction P-P $\rightarrow$ $\pi$ + X extremely well.(see
Fig. 1) He then asked me to think about the effect of adding thermal
fluctuations and with the help of Graham Frye and Edmond Schonberg at
Yeshiva University (and also some help from Mitchell Feigenbaum) we
found the correct covariant method of adding thermal fluctuations to
the hydrodynamic flow which is now called the Cooper-Frye-Schonberg
formula \cite{fred1}, \cite{fred2}. Based on these ideas we were able
to fit both the transverse momentum distribution as well as the
rapidity distribution extremely well (see Fig. 2).

\vspace{.4cm}
\epsfxsize=6cm
\epsfysize=6cm
\centerline{\epsfbox{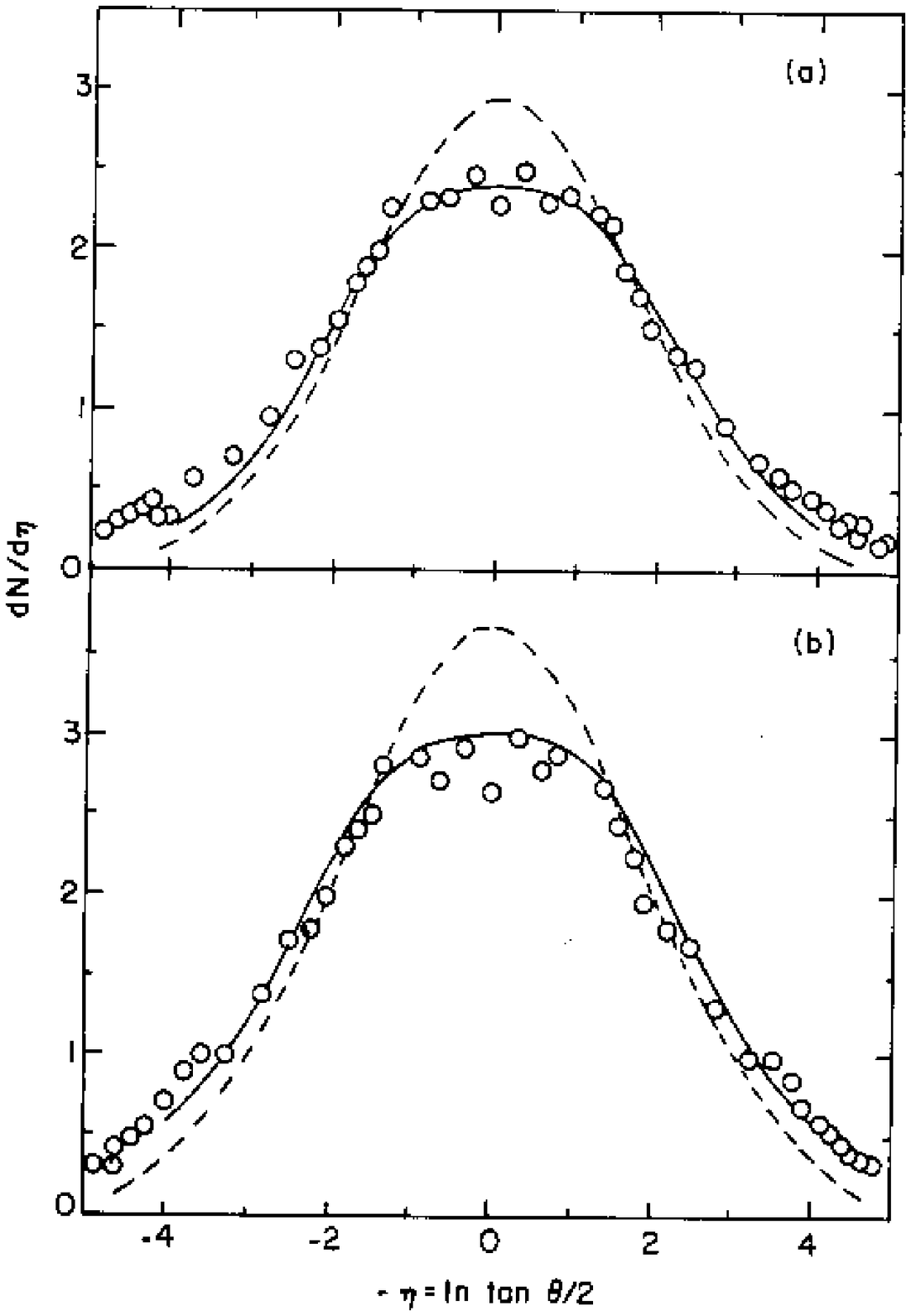}}
\vspace{.35cm}
{FIG. 1. {\small{Comparison of the experimental C.M.  rapidity
 distribution of outgoing pions from Proton Proton collisions at the
 ISR at (a) $p_{ISR}$ = $15.4$ GeV/c and (b) $p_{ISR}$ = $26.7$ GeV/c
 compared with the no adjustable parameter result of the Landau Model
 from \cite{pete2}}}}\\

The hydrodynamical model, which was considered ``heretical" when
applied to proton-proton collisions,as opposed to Feynman scaling
models, was resurrected by Bjorken \cite{bjorken1} in 1983 to describe
relativistic heavy ion collisions. The use of the hydrodynamical model
to study heavy-ion collisions is now a sophisticated mini-industry
using 3-dimensional fluid codes and sophisticated equations of state
based on lattice QCD. The effects of resonance decays have also been
recently included. A recent fit to single particle inclusive spectra
for Pb-Pb collisions at 160 AGeV and S-S collisions at 200AGev
\cite{schlei1} is shown in Fig. 3.

\vspace{.4cm}
\epsfxsize=8cm
\epsfysize=6cm
\centerline{\epsfbox{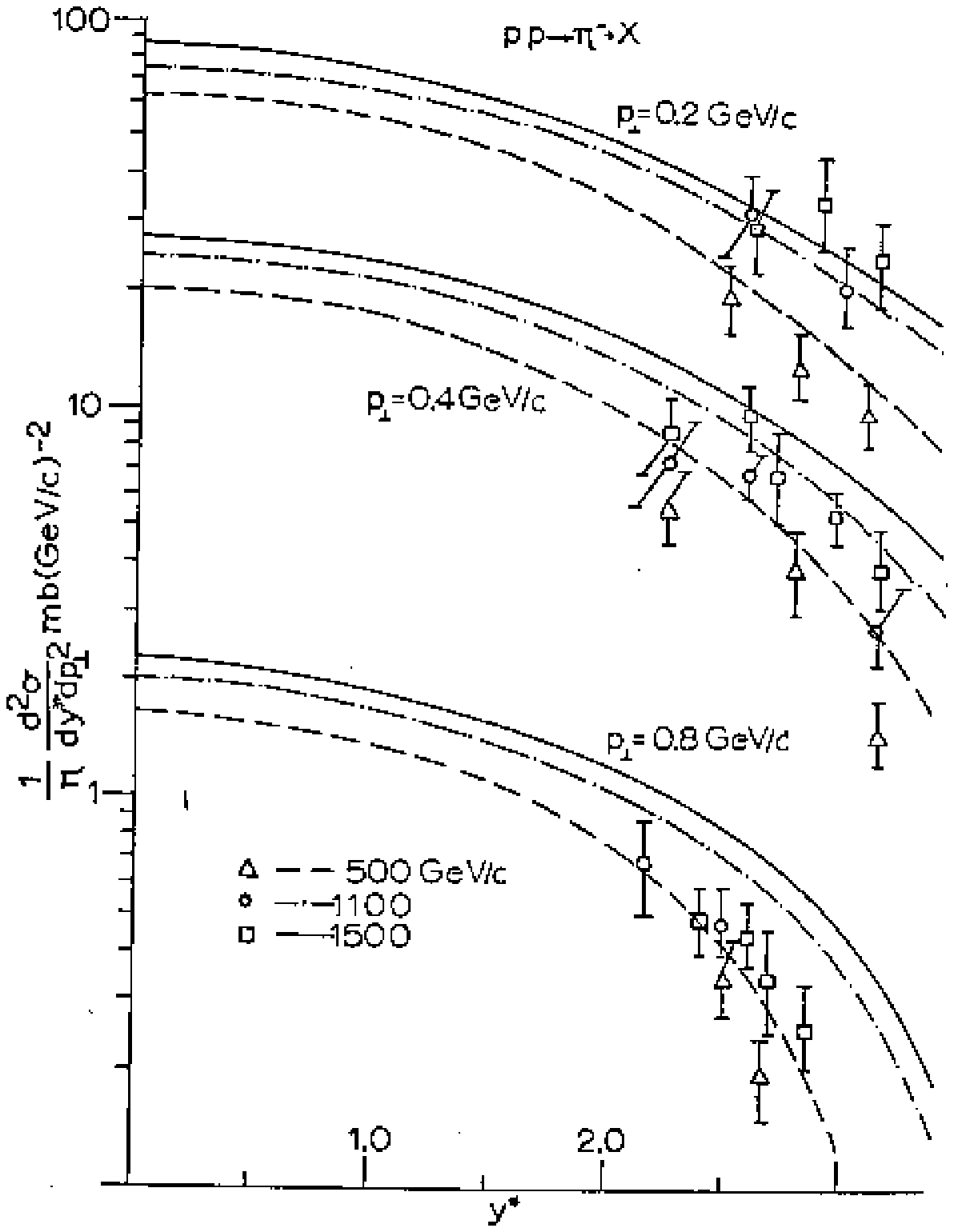}}
\vspace{.35cm}
{FIG. 2.{\small{Comparison of both transverse momentum and rapidity
distributions for an ISR experiment with the Cooper-Frye-Schonberg
reinterpretation of the Landau Model.  from \cite{fred3}}}}\\
 
  After the initial successes of the hydrodynamical model, Carruthers
and Zachariasen \cite{petefred1} \cite{petefred2} made the first
transport approach to particle production based on the covariant
Wigner-transport equation. This further work of Pete's inspired me and
my collaborators David Sharp \cite{sharp1} and Mitchell Feigenbaum
\cite{mitch1} to study particle production in $\lambda \phi^4 $in a
mean field approximation using the formalism espoused by Carruthers
and Zachariasen. Our efforts in the 70's were hampered by two factors:
first we were not sure what the appropriate field theory was, and
secondly computers were not large enough or fast enough to perform
accurate simulations of time evolution problems.  This covariant
transport theory approach turned out not to be the most convenient way
to study initial value problems and would later be replaced by a
non-covariant approach where the time was singled out.

\vspace{-1cm}
\begin{figure}
\hspace*{0.0cm}
\psfig{figure=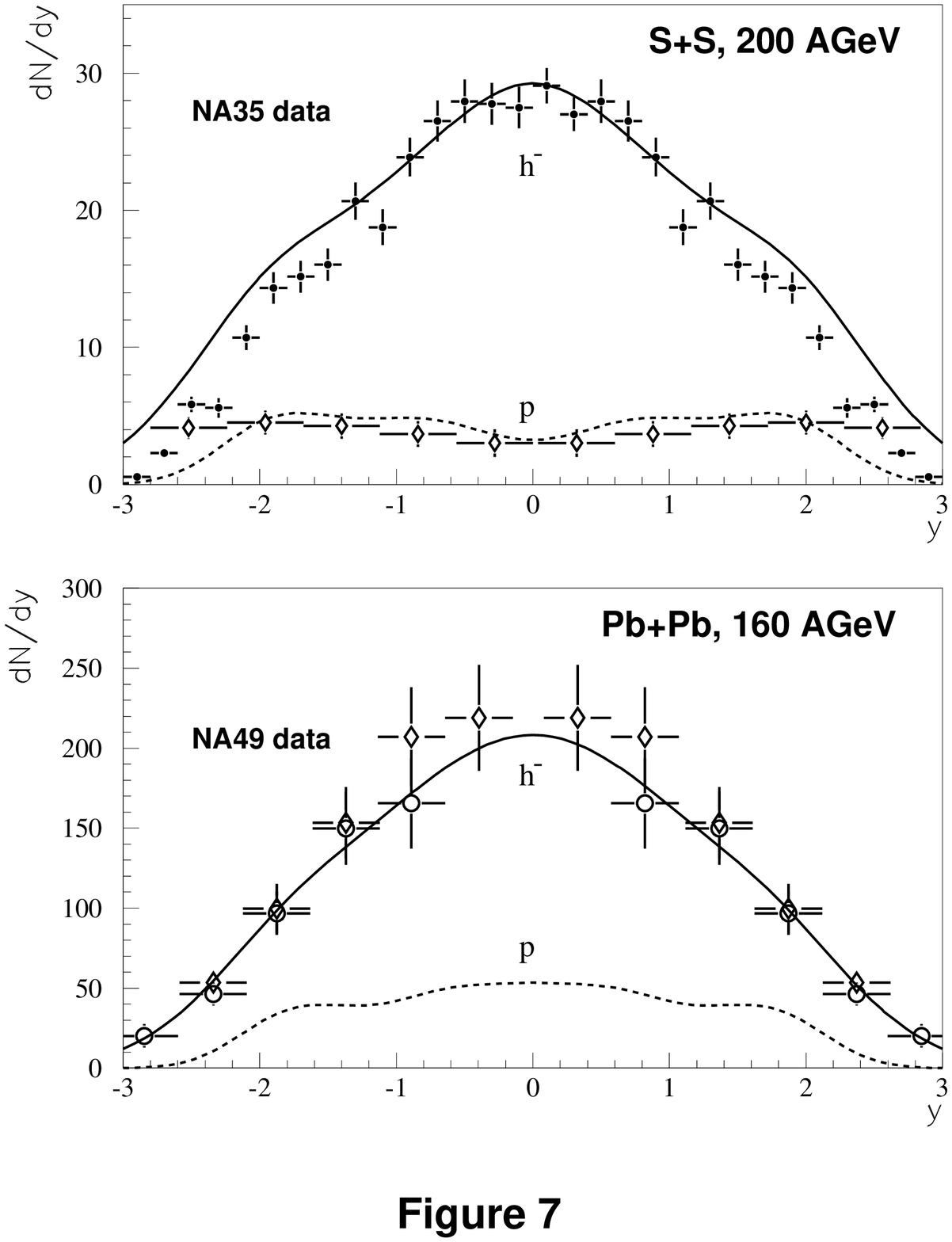,%
bbllx=0.0cm,bblly=3.5cm,bburx=20.0cm,bbury=29.7cm,%
width=7.0cm,clip=}
\end{figure}
\vspace{-.2cm}
{FIG. 3. {\small{Fit of recent Pb+Pb and S+S data of the NA49 and NA35
collaborations by Schlei et. al. \cite{schlei1}}}}\\

Starting in the mid-80's two new approaches were taken to understand
time evolution problems in field theory, both related in spirit to the
covariant transport approach of Carruthers and Zachariasen which
utilized a mean field approximation to close the coupled equations for
the distribution functions. The first method was to assume a Gaussian
ansatz in a time dependent variational method in the Schrodinger
picture, \cite{pi1} \cite{pi2}. The second method was to directly
study the time evolution of the Green's functions in the Heisenberg
picture in a leading order in large-$N$ approximation \cite{emil1}
\cite{emil2} \cite{emil3}.

Both these methods, which are mean-field approximations, lead to a
well posed initial value problem for the time evolution of a field
theory. In the mean-field approximation, we discovered that one has to
solve at least 10,000 equations for the Fourier modes of the quantum
field theory in order to be in the continuum limit (i.e. for the
coupling constant to run according to the continuum renormalization
group). Thus, although much of the formalism was worked out by me and
my collaborators in the mid and late 80's, \cite{formal} it was not
until the advent of parallel computation that numerical algorithms
were fast enough to make these calculations practical. The first
initial simulation attempts were presented in Santa Fe in 1990 at a
workshop on intermittency that Pete asked me to help organize
\cite{fred4}. These first simulations took weeks of dedicated machine
time. With the advent of the connection machine CM-5 at the ACL at Los
Alamos simulations can now be done in a few hours finally making
serious studies possible.
 
 In the past few years, we have been able to consider two aspects of
Relativistic Heavy Ion Collisions. The first aspect is connected with
the production of the quark gluon plasma. The model we used was the
popping of quark- antiquark pairs out of the vacuum due to the
presence of Strong Chromoelectric fields. The mechanism we used was
based on Schwinger's calculation of pair production from strong
Electric Fields \cite{schwinger1}. Using this model and assuming boost
invariant kinematics we were able to show that many of Landau's
assumptions were verified--namely that from the energy flow alone one
could determine the particle spectrum, and that the fluid rapidity
spectra was the same as the particle rapidity spectra when one is in a
scaling regime. Also the hydrodynamical prediction for the dependence
of the entropy density as a function of the proper time was verified
at long times. Numerical evidence for this will be displayed below.

The second problem we considered was the dynamics of a non-equilibrium
chiral phase transition. In this case because of the phase transition,
one can obtain single particle spectra which are different from a
local equilibrium flow such as that given by Landau's model. We found
that when the evolution proceeds through the spinodal regime, where
the effective mass becomes negative, low momentum modes grow
exponentially for short periods of proper time. This leads to an
enhancement of the low momentum spectrum over what would have been
found in an equilibrium evolution. While considering this second
problem we realized that the Cooper-Frye-Schonberg formula was valid
even in a field theory evolution, provided one interprets the field
theory interpolating number density as the single particle phase space
distribution function of a classical transport theory \cite{emil3}.

It is safe to say that much of my career was stimulated by Pete
encouraging me to understand multiparticle production in high energy
collisions and by him freely sharing all of his intuition about this
subject.  It has taken 20 years to show that our original thinking
back in 1973 was mostly correct!!

Landau's model \cite{landau1} of multiparticle production was based on
very few assumptions. He first made an assumption about the initial
condition of the fluid. Namely, he assumed that after a high energy
collision some substantial fraction ($\approx 1/2$) of the kinetic
energy in the center of mass frame was dumped into a Lorentz
contracted disc with transverse size that of the smallest initial
nuclei. He then assumed the flow of energy was describable by the
relativistic hydrodynamics of a perfect fluid having an
ultrarelativistic equation of state $p={\varepsilon \over 3}$.(This
assumption was later modified by later workers as knowledge first from
the bag model and then from lattice QCD became available).  The motion
of the fluid is described by a collective velocity field $
u^{\mu}(x,t)$. Once the collective variables, energy density, pressure
and collective velocity are postulated, along with the equation of
state, the equations of motion follow from the conservation of
energy-momentum.  The same assumption of a perfect relativistic fluid
is used in Cosmology. One solves the Einstein equations for the
curvature assuming an energy momentum tensor that is a perfect
fluid. (see \cite{Weinberg} )

 The rest of the dynamics is embodied in the initial and final
conditions. We stated already that the initial condition is to assume
that the initial energy density distribution is constant in a Lorentz
contracted pancake.  In our field theory calculations using strong
fields, we will assume that all the energy density is in the initial
semiclassical electric (chromoelectric) field, and equate the energy
density $E^2(x,t)$ with Landau's $\varepsilon(x,t)$. Thus we will not
attempt to derive this initial condition but will retain this
assumption.  However various event generators do find similar energy
densities to those obtained from a hydrodynamical scaling expansion.
The final boundary condition is to state that when the energy density
$\varepsilon$ reduces to a critical value $\varepsilon_c$ which is
approximately one physical pion/ pion compton wavelength$^3$ in a
comoving frame, then there are no more interactions and one then
calculates the spectrum at that ``freeze out" surface. The original
method of Landau to determine the particle distribution was to
identify the pion velocity with the collective velocity and assume
that the number of particles in a bin of particle rapidity was equal
to the energy in that bin divided by the energy of a single pion
having that rapidity.  An alternative was to assume the number
distribution was proportional to the entropy distribution. This ansatz
was later modified by myself and my collaborators by assuming there
was a local thermal distribution of pions in the comoving frame at
temperature $T_c$ described by
\begin{equation} g(x,k) =  g_{\pi}  \{ {\rm exp} [k^{\mu}u_{\mu}/T_c]
-1^{-1}. 
\end{equation}
The particle distribution was then given by: 
\begin{equation}
E {d^3N \over d^3k} = {d^3N \over \pi dk_{\perp}^2 dy} =
\int g(x,k) k^{\mu} d \sigma_{\mu}
\end{equation}
where $\sigma_{\mu}$ is the surface defined by
$\varepsilon=\varepsilon_c$.  We will verify that both ideas can be
justified by our field theory calculations.

Let us now let us look at the hydrodynamics of a perfect fluid.  In
the rest frame (comoving frame) of a perfect relativistic fluid the
stress tensor has the form:
\begin{equation}
{T_{\mu\nu} = {\rm diagonal}\; (\epsilon, p,
p, p)}
\end{equation} 
Boosting by the relativistic fluid velocity four vector $u^{\mu}(x,t)$
one has: 
\begin{equation}
{T_{\mu\nu} = (\varepsilon + p) u^{\mu}u^{\nu} - pg ^{\mu\nu}}
\end{equation}

From a hydrodynamical point of view, flat rapidity distributions seen
in multiparticle production in p-p as well as A-p and A-A collisions
are a result of the hydrodynamics being in a scaling regime for the
longitudinal flow. (More exact 3-D numerical simulations with
sophisticated equations of state have now been performed. The
interested reader can see for example \cite{schlei1}).

That is for $v={z/t}$ (no size scale in the longitudinal dimension)
the light cone variables $\tau, \eta$:
\begin{equation}
{z = \tau \sinh \eta; t=\tau
\cosh\eta} 
\end{equation}
become the fluid proper time $\tau=t (1-v^{2})^{1/2}$ and fluid rapidity:
\begin{equation}{\eta= 1/2 \ln [(t-x)/(t+x)] \Rightarrow 1/2 \ln
[(1-v)/(1+v)] = \alpha} 
\end{equation}
Letting $ u^{0} = \cosh \alpha ;u^{3} = \sinh \alpha$, we have when $v
=zt$ that $\eta = \alpha$, the fluid rapidity. If one has an effective
equation of state $p = p(\varepsilon)$ then one can formally define
temperature and entropy as follows: \begin{equation} {\varepsilon + p
=Ts; d\varepsilon = T ds; \ln s = \int d\varepsilon / (\varepsilon
+p)} \label{eq:s}
\end{equation}
 Then the equation: $$u^{\mu}\partial^{\nu}T_{\mu\nu}=0$$
becomes:
\begin{equation}
{\partial^{\nu} (s(\tau) u_{\nu})=0}
\end{equation}
Which in 1 + 1 dimensions becomes
\begin{equation}
{ds/d\tau + s/\tau =0\; {\rm or}\; s\tau = {\rm constant}}
\end{equation}

The assumption of Landau's hydrodynamical model is that the two
projectiles collide in the center of mass frame leaving a fixed
fraction (about 1/2) of their energy in a a Lorentz contracted disc
(with the leading particles going off).  The initial energy density
for the flow can then be related to the center of mass energy and the
volume of the Lorentz contracted disk of energy. It is also assumed
that the flow of energy is unaffected by the hadronization process and
that the fluid rapidity can be identified in the out regime with
particle rapidity.  Thus after hadronization the number of pions found
in a bin of fluid rapidity can be obtained from the energy in a bin of
rapidity by dividing by the energy of a single pion having that
rapidty. When the comoving energy density become of the order of
$\varepsilon_c$ we are in the out regime. This determines a surface
defined by
\begin{equation}
{\varepsilon_c(\tau_{f}) = m_{\pi}/V_{\pi}}
\end{equation}
On that surface of constant $\tau$,
\begin{eqnarray}
{dN \over d\eta} &=& {1 \over m_{\pi}u^{0}}{dE \over d\eta }= {1 \over
m_{\pi}\cosh\alpha}  \int T^{0\mu} {d\sigma_{\mu} \over d\eta }
\nonumber \\ 
  d\sigma_{\mu} &=& A_{\bot}(dz, -dt) = 4\pi
a^{2} \tau_{f}(\cosh\eta, -\sinh\eta) \nonumber \\
{dN \over d\eta}  &=& {A_{\bot} \over m_{\pi} \cosh \alpha} 
[(\varepsilon +p) \cosh \alpha \cosh(\eta -\alpha)- p\cosh
\eta]\nonumber \\ 
\label{eq:landenergy} 
\end{eqnarray}
where $A_{\bot}$ is the transverse size of the system at freezeout. 
If we are in the scaling regime where $\eta=\alpha$ then

\[{dN \over d\eta} ={A_{\bot} \over m_{\pi}} 
 \varepsilon(\tau_{f}). \] which is a flat distribution in fluid
rapidity. At finite energy, where scaling is not exact, only the
central region is flat and instead one gets a distribution which is
approximately Gaussian in rapidity. Exact numerical simulations (see
for example \cite{fred2}) show that the isoenergy curves do indeed
follow a constant $\tau$ curve in the central region, so that the
scaling result does apply for particle production in the central
rapidity region at high but finite center of mass energy.

In Landau's model one needed an extra assumption to identify the
collective fluid rapidity $\alpha$ with particle rapidity $y=1/2 \ln
[(E_{\pi} + p_{\pi}) /(E_{\pi}-p_{\pi})]$, where $p_{\pi}$ is the
longitudinal momentum of the pion. What results from our field theory
simulations of both the production of a fermion-anti fermion pairs
from strong Electric fields (the Schwinger mechanism
\cite{schwinger1}) as well as in the production of pions following a
chiral phase tranisition in the $\sigma$ model, is that if we make the
kinematical assumption that the quantum expectation values of
measurables are solely a function of $\tau$ we will obtain a flat
rapidity distribution for the distribution of particles. We can prove,
using a coordinate transformation, that the distribution of particles
in fluid rapidity is exactly the same as the distribution of particles
in particle rapidity.  Furthermore we also will find that it is a good
approximation to use eq. \ref{eq:landenergy} to determine the spectra
of particles.  For the full single particle distribution $ E {dN \over
d^3p}$ we will find that the Cooper-Frye formula is valid with the
identification of the interpolating number density with the single
particle distribution function of classical transport theory.

\section{Production and Time evolution of a quark-antiquark Plasma} 
Our model for the production of the quark-gluon plasma begins with the
creation of a flux tube containing a strong color electric field. If
the energy density of the chromoelectric field gets high enough (see
below) the quark-anti quark pairs can be popped out of the vacuum by
the Schwinger mechanism \cite{schwinger1}.  For simplicity, here we
discuss pair production (such as electron-positron pairs) from an
abelian Electric Field and the subsequent quantum back-reaction on the
Electric Field. The extension to quark anti-quark pairs produced from
a chromoelectric field is straightforward.  The physics of the problem
can be understood for constant electric fields as a simple tunneling
process. If the electric field can produce work of at least twice the
rest mass of the pair in one compton wavelength, then the vacuum is
unstable to tunnelling. This condition is: $$ eE {\hbar \over mc} \ge
2 m c^2 $$ which leads to a critical electric field of order $ {2 m^2
c^3 \over \hbar e} $

The problem of pair production from a constant Electric field
(ignoring the back reaction) was studied by J. Schwinger in 1951
\cite{schwinger1} . The WKB argument is as follows: One imagines an
electron bound by a potential well of order $|V_{0}|\approx 2m$ and
submitted to an additional electric potential $eEx$ .  The ionization
probability is proportional to the WKB barrier penetration factor:

$$\exp [-2 \int_{o}^{V_{o}/e} dx \lbrace 2m (V_{o} -|eE| x)\rbrace^{1/2}] 
= \exp (-{4 \over 3} m^{2}/|eE|)
$$
In his classic paper Schwinger was able to analytically solve for the
effective Action in a constant background electric field and determine
an exact pair production rate: 
$$w = [\alpha E^{2}/(2\pi^{2})] {\sum_{n=1}^{\infty}} {(-1)^{n+1}
\over n^{2}} \exp (-n\pi m^{2}/|eE| ). $$
By assuming this rate could be used when the Electric field was slowly
varying in time, the first back reaction calculations were attempted
using semi classical transport methods. Here we directly solve the
field equations in the large-N approximation \cite{emil2}. We assume
for simplicity that the kinematics of ultrarelativistic high energy
collisions results in a boost invariant dynamics in the longitudinal
($z$) direction (here $z$ corresponds to the axis of the initial
collision) so that all expectation values are functions of the proper
time $\tau = \sqrt{t^2-z^2}$.We introduce the light cone variables
$\tau$ and $\eta$, which will be identified later with fluid proper
time and rapidity . These coordinates are defined in terms of the
ordinary lab-frame Minkowski time $t$ and coordinate along the beam
direction $z$ by
\begin{equation}
       z= \tau \sinh \eta  \quad ,\quad t= \tau \cosh \eta \,.
\label{boost_tz.tau.eta}
\end{equation}
 The Minkowski line element in these coordinates has the form
\begin{equation}
{ds^2} = {- d \tau^2 + dx^2   +dy^2 +{\tau}^2 {d \eta}^2 }\,.
\label{boost_line_element}
\end{equation}
Hence the metric tensor is given by
\begin{equation}
 g_{\mu \nu} = {\rm diag} (-1, 1, 1, \tau^2).
\end{equation}
 
The QED  action in curvilinear coordinates is:
\begin{eqnarray}
S &&= \int d^{d + 1}x \, ({\rm{det}}\, V) [ -{i \over 2}
\bar {\Psi} \tilde{\gamma}^{\mu}
\nabla_{\mu} \Psi+ {\frac{i}{2}} (\nabla^{\dag}_{\mu}\bar {\Psi} )
\tilde{\gamma}^{\mu} \Psi  \nonumber \\
&& -i m \bar {\Psi}\Psi- {1 \over 4}F_{\mu \nu}F^{\mu \nu} ],
\label{boost_Sf}
\end{eqnarray}
where 
\begin {equation}
\nabla_{\mu} \Psi \equiv (\partial_{\mu} + \Gamma_{\mu} -ieA_{\mu})
\Psi 
\end{equation}

Varying the action leads to the Heisenberg field
equation:
\begin{equation}
\left( \tilde{\gamma}^{\mu}\nabla_{\mu} + m \right) \Psi=0\,,
\end{equation}

\begin{equation}
\left[ \gamma ^0 \left(\partial_\tau+{1 \over 2\tau}\right)
+{\bf \gamma}_\perp\cdot \partial_\perp
+ {\gamma^3 \over \tau}(\partial_\eta -ieA_\eta)+ m \right] \Psi =0\,,
\label{boost_Dirac}
\end{equation}
 and the Maxwell equation:
$E=E_z(\tau)= - \dot{A}_{\eta}(\tau)$
\begin{equation}
{1 \over \tau}  {dE(\tau) \over d\tau} =  {e \over 2} \left \langle
\left[ \bar{\Psi}, \tilde {\gamma}^{\eta} \Psi \right] \right \rangle
={e \over 2 \tau} \left \langle \left[ \Psi^{\dagger}, \gamma^0
\gamma^3 \Psi \right] \right \rangle .
\label{boost_MaxD2}
\end{equation}

\vspace{.2cm}
\epsfxsize=7cm
\epsfysize=5cm
\centerline{\epsfbox{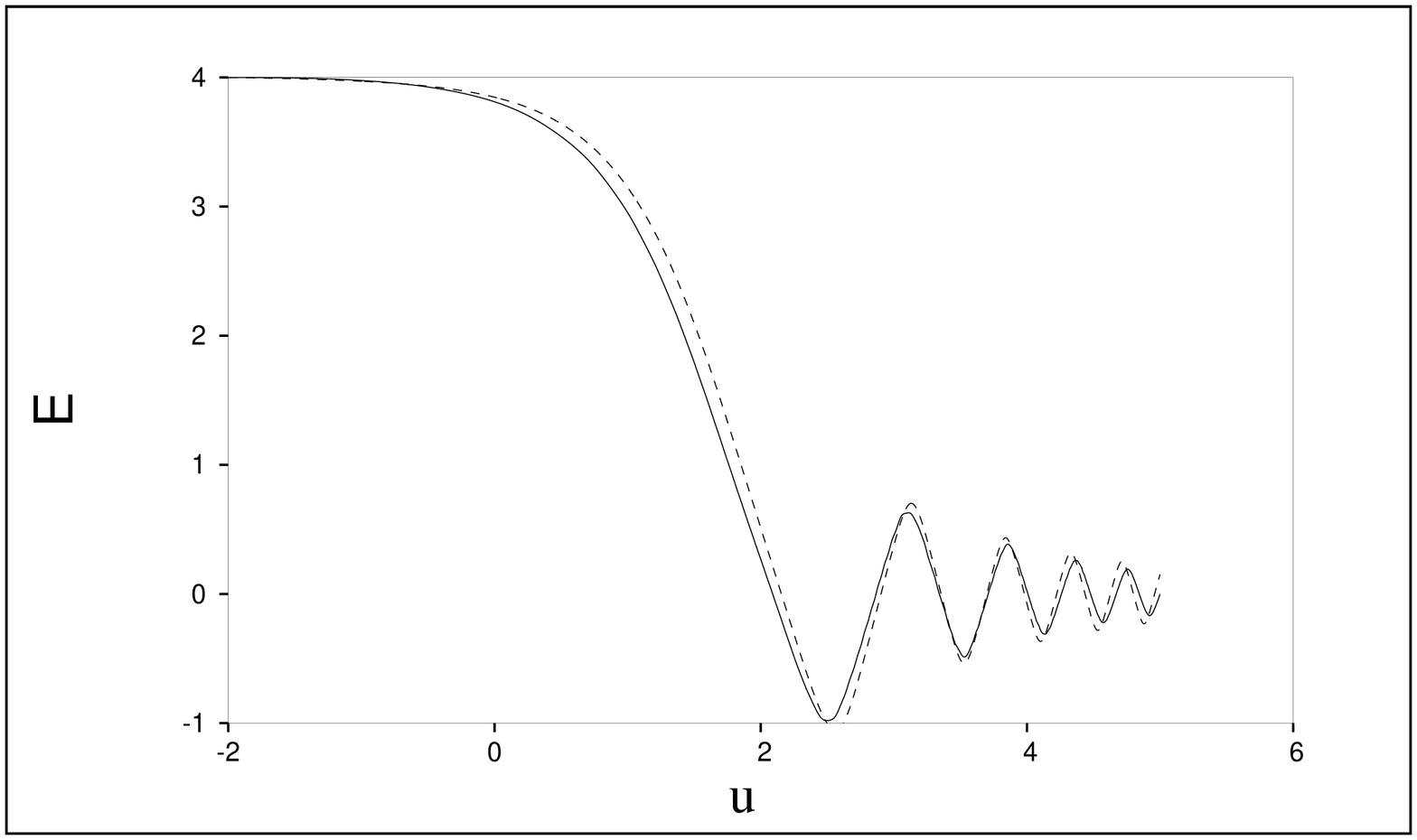}}
\epsfxsize=7cm
\epsfysize=5cm
\centerline{\epsfbox{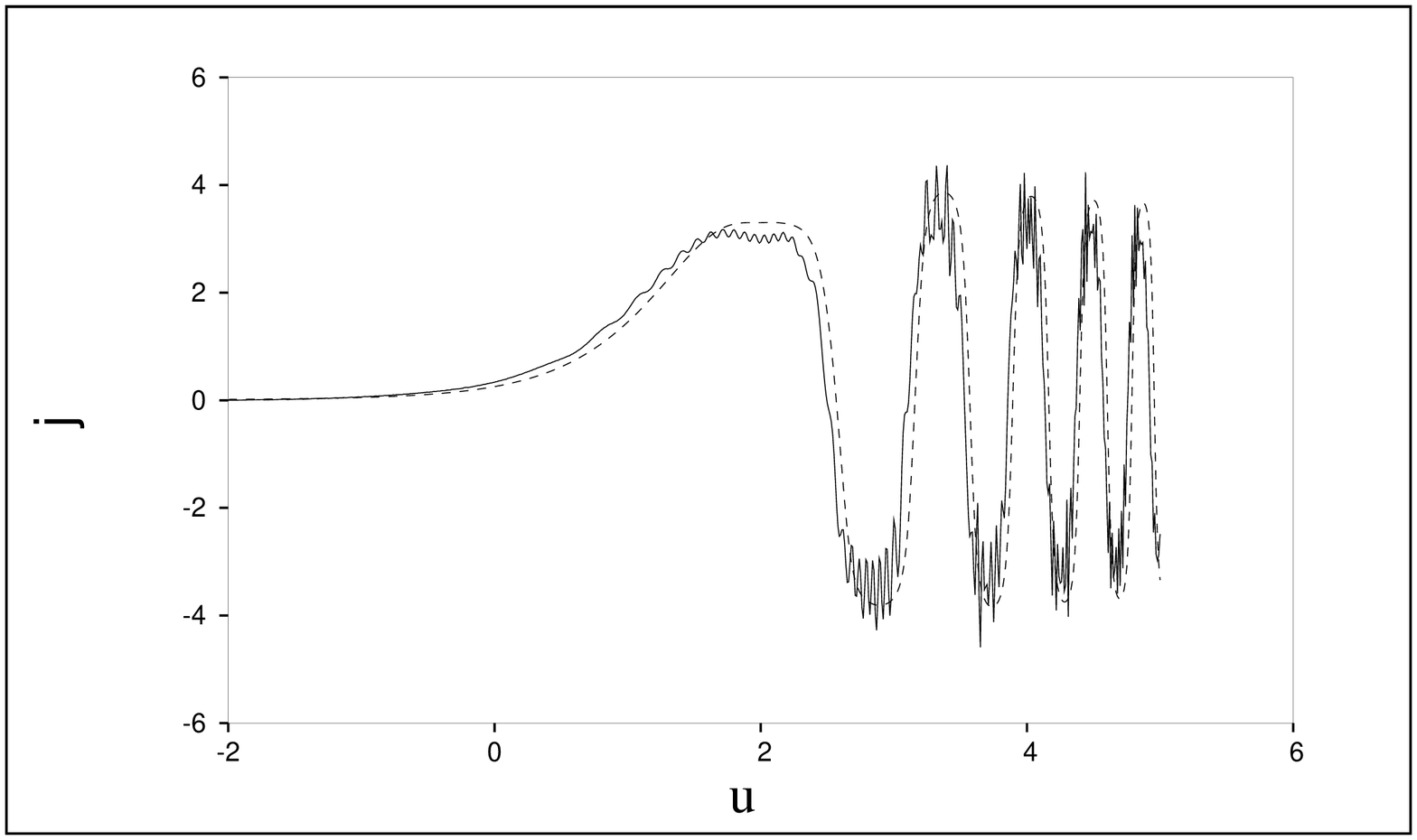}}
\vspace{.35cm}
{FIG. 4. {\small{Proper time evolution of $E$ and $j$ as a function of
$u=\ln ({\tau \over \tau_0})$ for an initial $E=4$.}}}\\  

We expand the fermion field in terms of Fourier modes at fixed proper
time: $\tau$, 
\begin{eqnarray}
\Psi (x) &=& \int [d{\bf k}] \sum_{s}[b_{s}({\bf k})
\psi^{+}_{{\bf k}s}(\tau)
 e^{i k \eta} e^{ i {\bf{p}} \cdot {\bf x}} \nonumber \\
&&+d_{s}^{\dagger}({\bf{-k}}) \psi^{-}_{{\bf{-k}}s}(\tau)
e^{-i k \eta} e^{ - i {\bf{p}} \cdot {\bf x}}  ].
\label{boost_fieldD}
\end{eqnarray}
The $\psi^{\pm}_{{\bf k}s}$ then obey
\begin{eqnarray}
 \left[\gamma^{0} \left({d\over d \tau}+{1 \over 2\tau}\right)
+ i \gamma_{\bf{{\perp}}}\cdot{\bf{k_{\perp}}}
+i {\gamma^{3}} \pi_{\eta}
 + m\right]\psi^{\pm}_{{\bf k}s}(\tau) = 0,
\label{boost_mode_eq_D}\end{eqnarray}
Squaring the Dirac equation:
\begin{equation}
\psi^{\pm}_{{\bf k}s} = \left[-\gamma^{0}\left( {d \over d\tau}
+{1 \over 2\tau}\right)
- i \gamma_{\bf{{\perp}}}\cdot {\bf{k_{\perp}}}
-i \gamma^{3} \pi_{\eta}+ m\right] \chi_{s} {f^{\pm}_{{\bf k}s} \over 
{\sqrt \tau}}
\, . \label{boost_psi_g}
\end{equation}
\begin{equation}
\gamma^{0}\gamma^{3}\chi_{s} = \lambda_{s} \chi_{s}\,
\end{equation}
with $\lambda_{s}=1$ for $s=1,2$ and $\lambda_{s}=-1$ for $s=3,4$,
we then get the mode equation:
\begin{equation}
\left( {d^2 \over d \tau^2}+
\omega_{\bf k}^2 -i \lambda_{s} \dot{\pi}_{\eta} \right )
f^{\pm}_{{\bf k}s}(\tau) = 0,
\label{boost_modef_D}
\end{equation}
\begin{equation}
 \omega_{\bf k}^2= \pi_{\eta}^2 +{\bf k}_{\perp} ^2 +m^2;~~ 
\pi_{\eta}={k_{\eta} -eA \over \tau}. \label{boost_omega_D}
\end{equation}

The back-reaction equation  in terms of the modes  is
\begin{equation}
{1 \over \tau}{dE(\tau) \over d\tau} = -{2e \over \tau^2}
\sum_{s=1}^{4}\int [d{\bf k}]
({\bf k}^2_{\perp} +m^2)
\lambda_{s}\vert f_{{\bf k}s}^{+}\vert ^2 ,
\label{boost_MaxD3}
\end{equation}
A typical proper time evolution of $E$ and $j$ is shown in
fig. 4. Here an initial value of $E=4$ was chosen.        
   
\subsection{Spectrum of Particles}

To determine the number of particles produced one needs to introduce
the adiabatic bases for the fields:
\begin{eqnarray}
\Psi (x) &&= \int [d{\bf k}] \sum_{s}[b_{s}^{0}({\bf k};\tau)
u_{{\bf k}s}(\tau)
 e^{-i \int \omega_{\bf{k}} d \tau} \nonumber \\
&&+d_{s}^{(0) \dagger}({\bf{-k}};\tau) v_{{\bf -k}s}(\tau)
 e^{i \int \omega_{\bf{k}}d \tau}] e^{i {\bf k \cdot x}}.
\label{adiab}
\end{eqnarray}

The operators $b_s ({\bf k})$ and $b_{s}^{(0)}({\bf k};\tau)$ are
related by a Bogolyubov transformation:
\begin{eqnarray}
b_{r}^{(0)}({\bf k};\tau) &=& \sum \alpha_{{\bf k} r}^s(\tau) b_s({\bf
k}) + \beta_{{\bf k} r}^s(\tau) d_s^{\dag}({\bf -k}) \nonumber \\
d_{r}^{(0)}({\bf -k};\tau) &=& \sum \beta_{{\bf k} r}^{*s}(\tau)
b_s({\bf k}) + \alpha_{{\bf k} r}^{*s}(\tau) d_s^{\dag}({\bf -k})
\end{eqnarray}
One finds that the interpolating phase space number density for the
number of particles (or antiparticles) present per unit phase space
volume at time $\tau$ is given by:
\begin{equation}
n({\bf k};\tau) = \sum_{r=1,2} \langle 0_{in} |b_{r}^{(0) \dag}({\bf
k};\tau) b_{r}^{(0)}({\bf k};\tau) |0_{in} \rangle = \sum_{s,r}
|\beta^s_{{\bf k} r}(\tau) |^2
\end{equation}

This is an adiabatic invariant of the Hamiltonian dynamics governing
the time evolution of the one and two point functions, and is
therefore the logical choice as the particle number operator. At
$\tau=\tau_0$ it is equal to our initial number operator. If at later
times one reaches the out regime because of the decrease in energy
density due to expansion it becomes the usual out state phase space
number density. Although this does not happen for the above pair
production in the Mean field approximation,(because we have not
allowed the electric field to dissipate due to the production of real
photons), reaching an out regime does happen in the $\sigma$ model if
the energy density decreases as a result of an expansion into the
vacuum.

The phase space distribution of particles (or antipartcles) in light
cone variables is
\begin{equation}
n_{\bf k}(\tau) =f(k_{\eta}, k_{\perp},\tau) = {d^6 N \over \pi^2
dx_{\perp}^2 dk_{\perp}^2d\eta dk_{\eta}}.
\label{interplc}
\end{equation}
A typical spectrum is shown in fig. 5 which shows the effect of the
Pauli- exclusion principle. The raw results and also the results of
averaging over typical experimental momentum bins are shown. This
latter result compares well with a transport approach including
Pauli-blocking effects (see \cite{emil2}).
\epsfxsize=8cm
\epsfysize=7cm
\centerline{\epsfbox{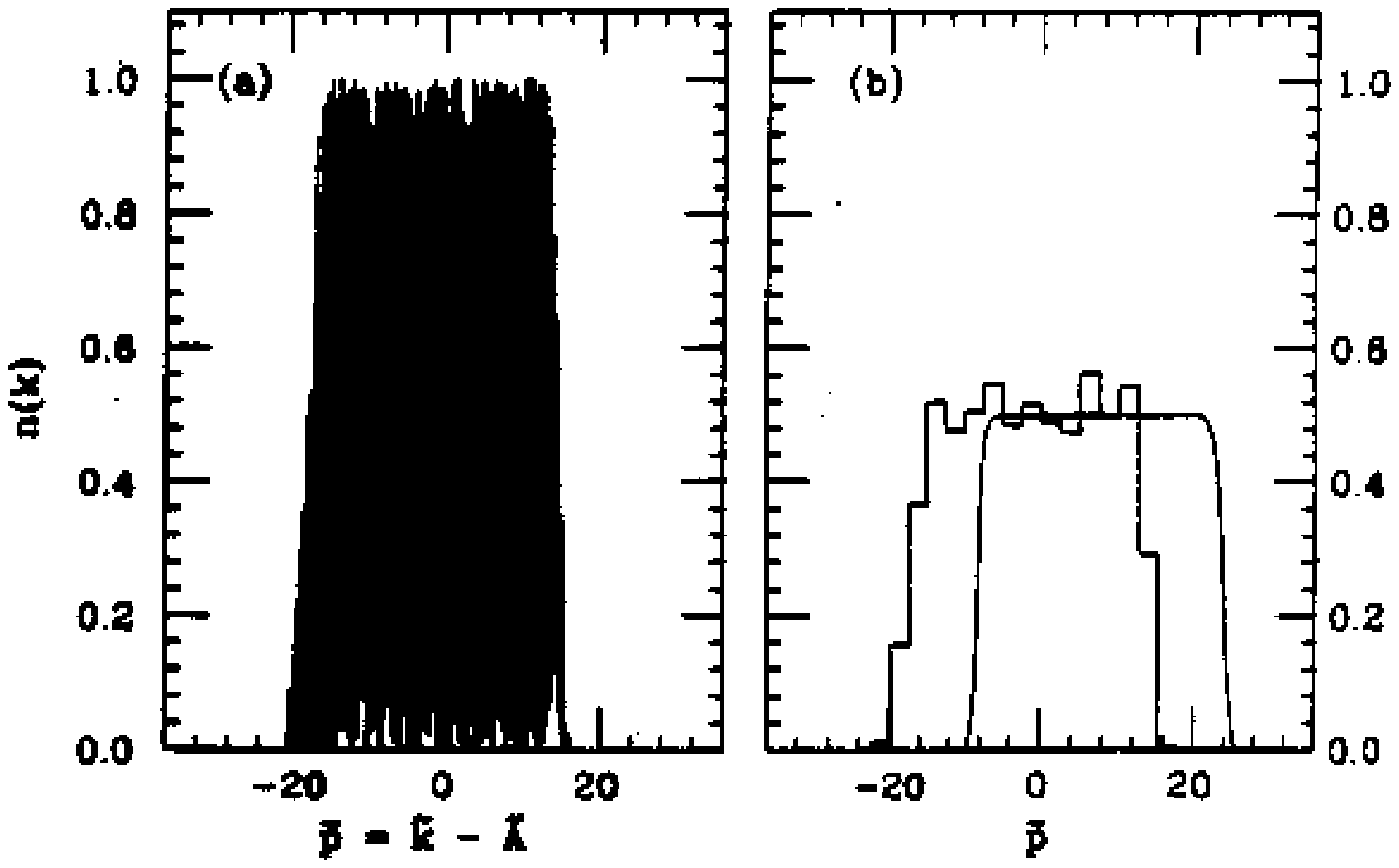}}
\vspace{0cm}
{FIG. 5. {\small{Comoving spectra of fermion pairs, before and after
binning for an initial electric field $E=4$ at $\tau$=$400$}}}\\
 
We now need to relate this quantity to the spectra of electrons and
positrons produced by the strong electric field (the production of
electrons and positrons from a strong electric field is our prototype
model for the production of the quark gluon plasma from strong
chromoelectric fields).  We introduce the particle rapidity $y$ and
$m_{\perp}= \sqrt{k_{\perp}^2 + m^2}$ defined by the particle
4-momentum in the center of mass coordinate system
\begin{equation}
k_{\mu} = (m_{\perp} \cosh y, k_{\perp},m_{\perp} \sinh y)
\label{kmu}
\end{equation}
The boost that takes one from the center of mass coordinates to the
comoving frame where the energy momentum tensor is diagonal is given
by $\tanh \eta= v = z/t$, so that one can define the ``fluid"
4-velocity in the center of mass frame as
\begin{equation}
u^{\mu} = (\cosh \eta,0,0, \sinh \eta)
\end{equation}
We then find that the variable 
\begin{equation}
\omega_k = \sqrt{m_{\perp}^2+{ k_{\eta}^2 \over \tau^2} } \equiv
k^{\mu}u_{\mu}
\end{equation}
has the meaning of the energy of the particle in the comoving frame.
The momenta $k_{\eta}$ that enters into the adiabatic phase space
number density is one of two momenta canonical to the variables
defined by the coordinate transformation to light cone
variables. Namely the variables
\begin{eqnarray}
 \tau = (t^2-z^2)^{1/2}  \qquad \,
\eta = \frac{1}{2} \ln \left({{t+z} \over{t-z}} \right) \nonumber
\end{eqnarray}
have as their canonical momenta
\begin{eqnarray}
k_{\tau}= Et/ \tau -k_z z/ \tau \quad \, k_{\eta}  = -Ez + t k_z.
\label{boost_4trans}
\end{eqnarray}
To show this we consider the metric $ ds^2= d\tau^2 - \tau^2 d \eta^2$
and the free Lagrangian
\begin{equation}
L = {m \over 2} g_{\mu \nu} {dx^{\mu} \over ds} {dx^{\nu} \over ds}
\end{equation}
Then we obtain for example
\begin{eqnarray}
k_{\tau} &=& m {d \tau \over ds} = m [({\partial \tau \over \partial
t})_z {d t
\over ds}+ ({\partial \tau \over \partial z})_t {d z \over ds}]
\nonumber \\
&=& {Et -k_z z  \over \tau} = k^{\mu} u_{\mu}
\end{eqnarray} 
The interpolating phase-space density $f$ of particles depends on
$k_{\eta}$, ${\bf k}_{\perp}$, $\tau$, and is $\eta$-independent.  In
order to obtain the physical particle rapidity and transverse momentum
distribution, we change variables from $(\eta, k_{\eta})$ to $(z, y)$
at a fixed $\tau$ where $y$ is the particle rapidity defined by
(\ref{kmu}). We have
\begin{equation}
E{d^3 N \over d^3 k} =
{d^3N \over  \pi dy\,dk_{\perp}^2 } = 
\int\pi dz~ dx_{\perp}^2 ~ J ~ f(k_{\eta},
k_{\perp},\tau)
 \label{boost_J}
\end{equation}
where the Jacobian $J$ is evaluated at a fixed proper time $\tau$
\begin{eqnarray}
 J  &=& \left| \matrix{
 {\partial k_{\eta}}/{\partial y}&{\partial
k_{\eta}}/{\partial z} \cr
{\partial \eta}/{\partial y}&
{\partial \eta}/{\partial z} \cr}\right|
= \left| \frac {\partial k_{\eta}}{\partial y}\frac{\partial\eta}
{\partial z} \right|\nonumber\\
&=&{ m_{\perp} \cosh(\eta-y) \over
\cosh \eta}={\partial k_{\eta}\over \partial z}|_{\tau} .
\label{boost_Jinv}
\end{eqnarray}
We also have 
\begin{eqnarray}
k_{\tau} = m_{\perp} \cosh(\eta-y); \quad
k_{\eta} = -\tau m_{\perp} \sinh(\eta-y)\,.
\label{boost_ptaueta}
\end{eqnarray}
 Calling the integration over the transverse dimension the effective
transverse size of the colliding ions $A_{\perp}$ we then obtain that:
\begin{equation}
{d^3N \over  \pi dy\,dk_{\perp}^2} = A_{\perp} \int dk_{\eta} 
f(k_{\eta}, k_{\perp},\tau) \equiv {d^3N \over  \pi d\eta\,dk_{\perp}^2}
\end{equation}
This quantity is independent of $y$ which is a consequence of the
assumed boost invariance. Note that we have proven using the property
of the Jacobean, that the distribution of particles in partical
rapidity is the same as the distribution of particles in fluid
rapidity!! verifying that in the boost-invariant regime that Landau's
intuition was correct.

We now want to make contact with the Cooper- Frye Formalism.  First we
note that the interpolating number density depends on $k_{\eta}$ and
$k_{\perp}$ only through the combination:
\begin{equation}
\omega_k = \sqrt{m_{\perp}^2+{ k_{\eta}^2 \over \tau^2} } \equiv
k^{\mu}u_{\mu}
\end{equation}
Thus $f(k_{\eta},k_{\perp}) = f(k_{\mu} u^{\mu})$ and so it depends on
exactly the same variable as the comoving thermal distribution! We
also have that a constant $\tau$ surface (which is the freeze out
surface of Landau) is parametrized as:
\begin{equation}
d\sigma^{\mu} = A_{\perp} (dz,0,0,dt) 
= A_{\perp}d\eta (\cosh \eta ,0,0,\sinh \eta )
\end{equation}
We therefore find
\begin{equation}
k^{\mu} d\sigma_{\mu} =  A_{\perp} m_{\perp} \tau \cosh(\eta-y)=
A_{\perp} |
dk_{\eta} |
\end{equation}
Thus we can rewrite our expression for the field theory particle
spectra as
\begin{equation}
{d^3N \over  \pi dy\,dk_{\perp}^2} = A_{\perp} \int dk_{\eta} 
f(k_{\eta}, k_{\perp},\tau)= \int f(k^{\mu}u_{\mu},\tau) k^{\mu}
d\sigma_{\mu}
\end{equation}
where in the second integration we keep $y$ and $\tau$ fixed.  Thus
with the replacement of the thermal single particle distribution by
the interpolating number operator, we get via the coordinate
transformation to the center of mass frame the Cooper-Frye formula.

Schwinger's pair production mechanism leads to an Energy Momentum
tensor which is diagonal in the($\tau,\eta, x_{\bot}$) coordinate
system which is thus a comoving one.  In that system one has:
\begin{equation}
{T^{\mu\nu} = {\rm diagonal}\; \lbrace \varepsilon(\tau),
p_{\parallel}(\tau), p_{\bot}(\tau), p_{\bot}(\tau) \rbrace}
\end{equation}
We thus find in this approximation that there are two separate
pressures, one in the longitudinal direction and one in the transverse
direction which is quite different from the thermal equilibrium
case. However only the longitudinal pressure enters into the
``entropy'' equation.

Only the longitudinal pressure enters into the ``entropy" equation
\begin{equation} 
\varepsilon + p_{\parallel} = Ts \label{eq:entro1}
\end{equation}
\[ {d(\varepsilon \tau)\over d \tau} + p_{\parallel} = E j_{\eta} \]
\[ {d(s\tau) \over  d \tau}= { E j_{eta} \over T} \]
In the out regime we  find as in the Landau Model
\[ s \tau = constant \]
as is seen in fig.6.

\epsfxsize=7cm
\epsfysize=6cm
\centerline{\epsfbox{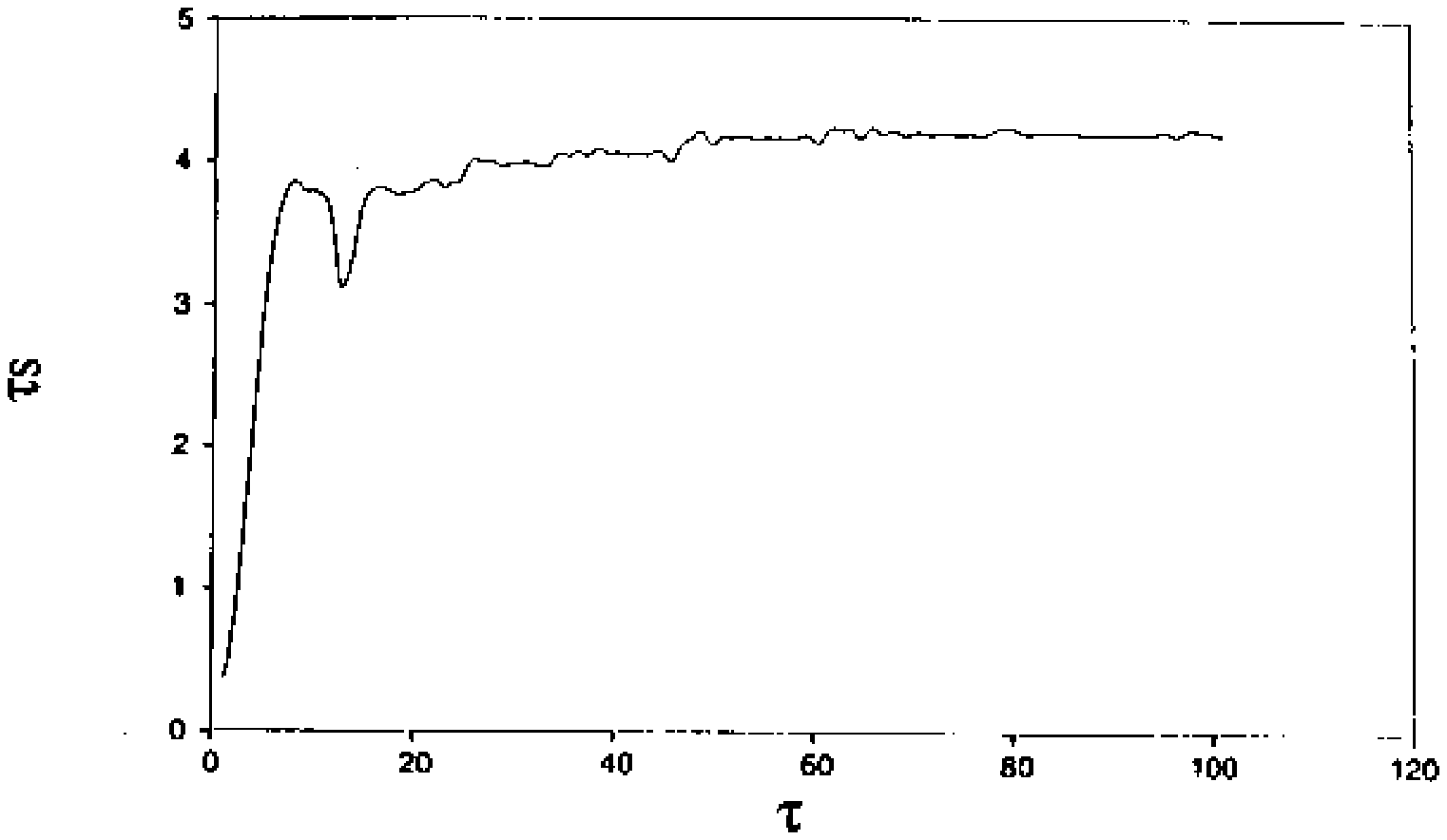}}
\vspace{.35cm}
{FIG. 6. {\small{Time evolution of $s \tau$ as a function of
$\tau$}}}\\

Here we have used \ref{eq:entro1} and the thermodynamic relation:
\[ d \varepsilon = T d s \]
to calculate the entropy from the energy density and longitudinal
pressure.  An alternative effective entropy can be determined from the
diagonal part of the full density matrix in the adiabatic number
basis.  The energy density as a function of proper time is shown in
fig.7.

\epsfxsize=7cm
\epsfysize=6cm
\centerline{\epsfbox{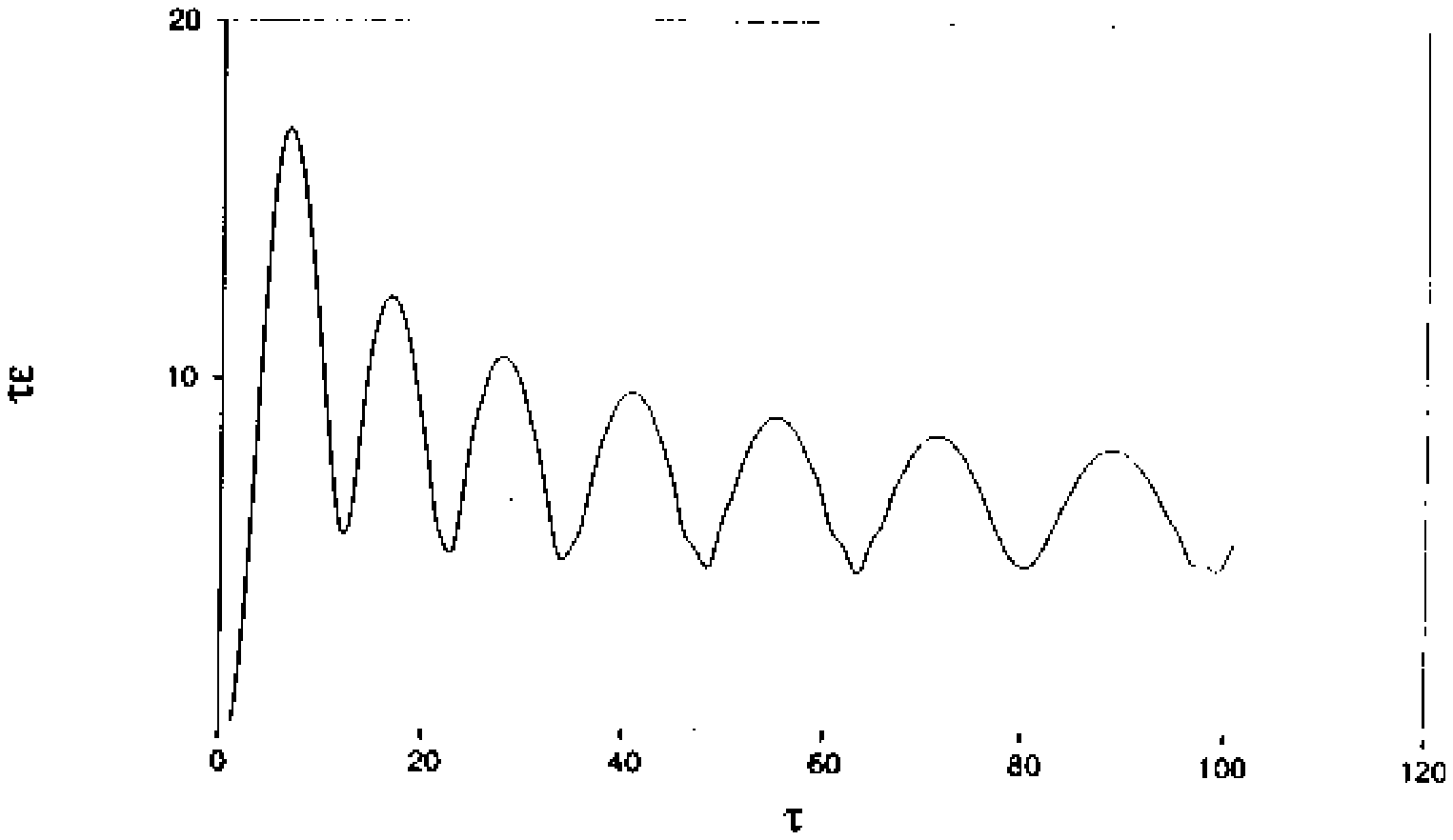}}
\vspace{.35cm}
{FIG. 7. {\small{Time evolution of $\varepsilon \tau$ as a function of
 $\tau$}}}\\

 For our one-dimensional boost invariant  flow we
find that the energy in a bin of fluid rapidity is just:

\begin{equation}
{dE \over d\eta} = \int T^{0\mu} d\sigma_{\mu} =
 A_{\bot} \tau \cosh \eta \varepsilon (\tau)
\end{equation}
which is just the (1 + 1) dimensional hydrodynamical result.Here
however $\varepsilon$ is obtained by solving the field theory equation
rather than using an ultrarelativistic equation of state.  This result
does not depend on any assumptions of thermalization.  We can ask if
we can directly calculate the particle rapidity distribution from the
ansatz:
\begin{equation}
{dN \over d \eta} = {1 \over m \cosh \eta} {dE \over d \eta} = { A_{\bot}
\over m} \varepsilon (\tau) \tau.
\end{equation}
We see from fig. 8. that this works well even in our case where we
have ignored rescattering, so that one does not have an equilibrium
equation of state.

\vspace{.3cm}
\epsfxsize=7cm
\epsfysize=5cm
\centerline{\epsfbox{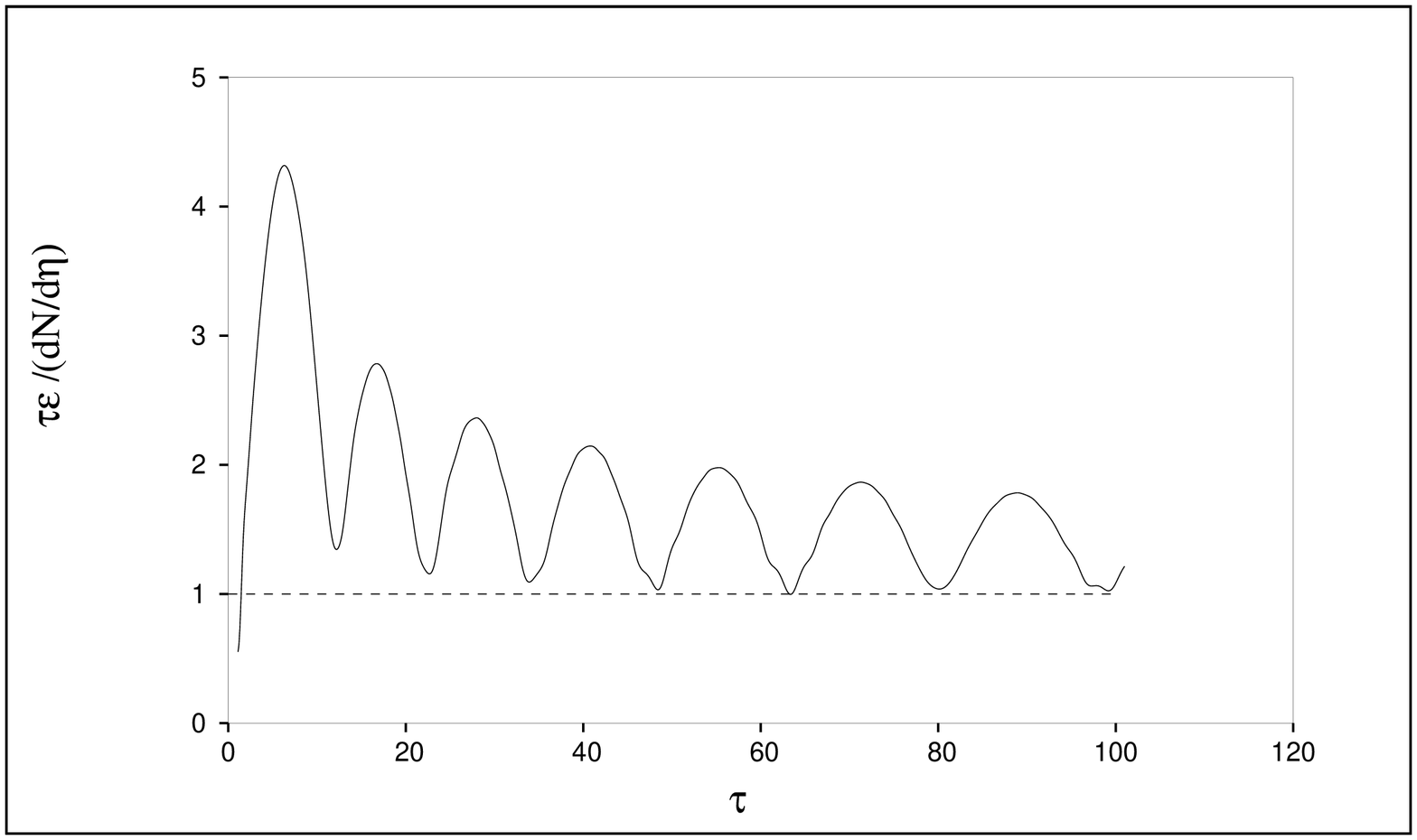}}
\vspace{.35cm}
{FIG. 8. {\small{The ratio of the approximate rapidity distribution
(${E \over m \cosh y} {dN \over dy}$) and exact rapidity distribution
as a function of $\tau$.}}}\\

 In the field theory calculation the expectation value of the stress
tensor must be renormalized since the electric field undergoes charge
renormalization.  We can determine the two pressures and the energy
density as a function of $\tau$.  Explicitly we have in the fermion
case. \[{\varepsilon (\tau) = < T_{\tau\tau} > = \tau \Sigma_{s} \int
[dk] R_{\tau\tau}(k) + E_{R}^{2}/2} \] where

\begin{eqnarray}
&& R_{\tau\tau}(k) =2(p_{\bot}^{2} 
+ m^{2}) (g_{0}^{+}| f^{+}|^{2} - g_{0}^{-}| f^{-}|^{2}) - \omega
\nonumber \\ 
&&- (p_{\bot}^{2} + m^{2}) (\pi + e \dot{A})^{2}/( 8
\omega^{5} \tau^{2})\nonumber \\
&& p_{\parallel} (\tau) \tau^{2} = < T_{\eta \eta} > = \tau \Sigma_{s}
\int [dk] 
\lambda_{s}\pi R_{\eta\eta}(k) - {1\over 2} E_{R}^{2} \tau^{2} 
\end{eqnarray}
where
\begin{eqnarray}
R_{\eta\eta} (k) && =2 | f^{+}|^{2} -
(2\omega)^{-1}(\omega+\lambda_{s}\pi)^{-1}  
 - \lambda_{s} e\dot{A}
/8\omega^{5}\tau^{2} \nonumber \\
&& -\lambda_{s} e\dot{E}/8\omega^{5}
-\lambda_{s}\pi/4\omega^{5}\tau^{2} + 5\pi\lambda_{s} (\pi + e
\dot{A})^{2}/(16 \omega^{7}\tau^{2}) \nonumber \\
\end{eqnarray}
and
\begin{eqnarray}
&& p_{\bot}(\tau)= < T_{yy} > = <T_{xx} > \nonumber \\
&& =(4 \tau)^{-1}
\sum_{s} \int [dk] \lbrace p_{\bot}^{2}(p_{\bot}^{2} +m^{2})^{-1}
R_{\tau\tau} -2\lambda\pi p_{\bot}^{2} R_{\eta\eta}\rbrace \nonumber \\
&&+ E_{R}^{2}/2.
\end{eqnarray}
Thus we are able to numerically determine the dynamical equation of
 state $p_{i}=p_{i}(\varepsilon)$ as a function of $\tau$.  A typical
 result is shown in fig. 9.

\vspace{.5cm}
\epsfxsize=7cm
\epsfysize=5cm
\centerline{\epsfbox{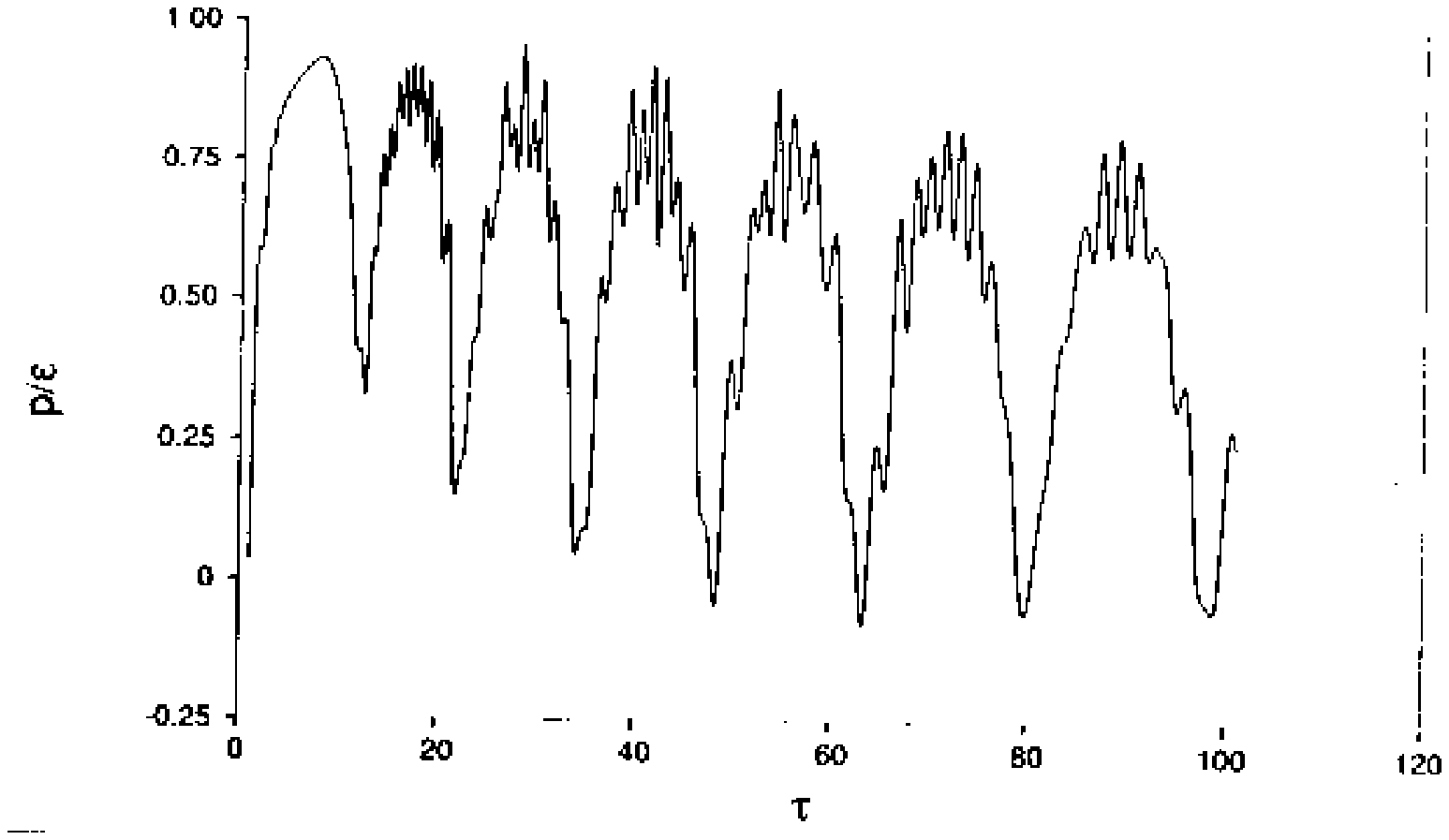}}
\vspace{.4cm}
{FIG. 9. {\small{Proper time evolution of ${p \over \varepsilon}$}}}\\

\section{Dynamical evolution of a non-equilibrium Chiral Phase Transition}

Recently there has been a growing interest in the possibility of
producing disoriented chiral condensates (DCC's) in a high energy
collision  \cite{Anselm,BjorkenIJMP,RajaWil}. This
idea was first proposed to explain CENTAURO events in cosmic ray
emperiments where there was a deficit of neutral pions
\cite{Cosmicrays}. It was proposed that a nonequilibrium chiral phase
transition such as a quench might lead to regions of DCC
\cite{RajaWil}.  To see whether these ideas made sense we studied
numerically \cite{DCC}the time evolution of pions produced following a
heavy ion collision using the linear sigma model, starting from the
unbroken phase.  The quenching (if present) in this model is due to
the expansion of the initial Lorentz contracted energy density by free
expansion into vacuum.  Starting from an approximate equilibrium
configuration at an initial proper time $\tau$ in the disordered phase
we studied the transition to the ordered broken symmetry phase as the
system expanded and cooled. We determined the proper time evolution of
the effective pion mass, the order parameter $<\sigma>$ as well as the
pion two point correlation function.  We studied the phase space of
initial conditions that lead to instabilities (exponentially growing
long wave length modes) which can lead to disoriented chiral
condensates.

 We showed that the expansion into the vacuum of the initial energy
distribution led to rapid cooling. This caused the system, initially
in quasi local thermal equilibrium to progress from the unbroken
chiral symmetry phase to the broken symmetry phase vacuum.  This
expansion is accompanied by the exponential growth of low momentum
modes for short periods of proper time for a range of initial
conditions.  This exponential growth of long wave length modes is the
mechanism for the production of disordered chiral condensates. Thus
the production of DCC's results in an enhancement of particle
production in the low momentum domain. Whether such an instability
occurs depends on the size of the initial fluctuation from the initial
thermal distribution.  The relevant momenta for which this exponential
growth occurs are the transverse momenta and the momenta $k_\eta =
-Ez+tp$ conjugate to the fluid rapidity variable $\eta =
\tanh^{-1}(z/t)$. We also found that the distribution of particles in
these momenta had more length scales than found in local thermal
equilibrium. When there is local thermal equilibrium, the length
scales are the mass of the pion and the temperature which is related
to the changing energy density, both of which depend on the proper
time $\tau$.

When we reexpress the number density in the comoving frame in terms of
the physically measurable transverse distribution of particles in the
collision center of mass frame, we find that there is a noticable
distortion of the transverse spectrum, namely an enhancement of
particles at low transverse momentum, when compared to a local
equilibrium evolution. We will consider two cases, one in which there
is exponential growth of low momentum modes due to the effective pion
mass going negative during the expansion, and one where the initial
fluctuations do not lead to this exponential growth. Both situations
will be compared to a purely hydrodynamical boost invariant
calculation based on local thermal equilibrium.  In determining the
actual spectra of secondaries, we find that the adiabatic number
operator of our large-$N$ calculation replaces the relativistic phase
space density $g(x,p)$ of classical transport theory in determining
the distribution of pariticles in rapidity and transverse momentum.
This makes it easy to compare our results with the hydrodynamical
calculation in the boost invariant approximation which assumes the
final pions are in local thermal equilibrium in the comoving frame.
 
  The model we use to discuss the chiral phase transition is the
linear sigma model described by the Lagrangian:
\begin{equation}
L= {1\over 2} \partial\Phi \cdot \partial\Phi - {1\over 4} 
\lambda (\Phi \cdot \Phi - v^2)^2 + H\sigma.
\end{equation}
The mesons  form an $O(4)$ vector $\Phi = (\sigma, \pi_i)$
This can be written in an alternative form by introducing the composite
field:
$\chi = \lambda (\Phi \cdot \Phi-v^2)$.
\begin{equation}
L_2 = -{ 1 \over 2} \phi_i (\Box + \chi) \phi_i + {\chi^2 \over 4
\lambda} + 
{1 \over 2} \chi v^2 + H \sigma
\end{equation}
The effective action to leading order in large $N$ is given by
\cite{DCC}
\begin{equation}
\Gamma[\Phi,\chi] = \int d^4x[ L_2(\Phi,\chi,H) + { i \over 2}N {\rm
tr~ln}
G_0^{-1}]
\end{equation}
$$
G_0^{-1}(x,y) = i[\Box + \chi(x)] ~\delta^4(x-y)
$$
Varying the action we obtain:
\begin{equation}
[\Box + \chi(x)] \pi_i = 0 ~~~~ [\Box + \chi(x)]\sigma = H,
\end{equation}
where here and in what follows, $\pi_i$,$\sigma$ and $\chi$ refer to
expectation values.  Varying the action we obtain
\begin{equation}
\chi= - \lambda v^2 + \lambda (\sigma^2 + \pi \cdot \pi) + \lambda
N  G_0 (x,x). 
\end{equation}
If we assume boost invariant kinematics \cite{fred2} \cite{bjorken1}
which result in flat rapidity distributions, then the expectation
value of the energy density is only a function of the proper time.
The natural coordinates for boost invariant ($v=z/t$) hydrodynamical
flow are the fluid proper time $\tau$ and the fluid rapidity $\eta$
defined as $$\tau\equiv(t^2-z^2)^{1/2}, \qquad \eta\equiv{1\over 2}
\log({{t-z}\over{t+z}}).
$$ To implement boost invariance we assume that mean (expectation)
values of the fields $\Phi$ and $\chi$ are functions of $\tau$ only.
We then get the equations:
\begin{eqnarray}
&&\tau^{-1}\partial_\tau\ \tau\partial_\tau\ \Phi_i(\tau)
 +\ \chi(\tau)\ \Phi_i(\tau) =
 H \delta_{i1} \nonumber \\
&&\chi(\tau) =\lambda\bigl(-v^2+\Phi_i^2(\tau)+
N G_0(x,x;\tau,\tau)\bigr),   
\end{eqnarray}
To determine the Green's function $G_0(x,y;\tau,\tau^{\prime})$ we
introduce the auxiliary quantum field $\phi(x,\tau)$ which obeys the
sourceless equation:
\begin{equation}
\Bigl(\tau^{-1}\partial_\tau\ \tau\partial_\tau\ -
\tau^{-2}\partial^2_\eta  
-\partial^2_\perp + \chi(x)\Bigr)
\phi(x,\tau)=0.
\end{equation}
$$
    G_0 (x,y;\tau, \tau^{\prime}) \equiv <T\{ \phi(x,\tau)
~\phi(y,\tau^{\prime})\}>. 
$$

We expand the quantum fields in an orthonormal basis: 
$$
\phi(\eta,x_\perp,\tau)\equiv{1\over{\tau^{1/2}}} \int \ddk\bigl(\exp(ik
x)
f_\kk(\tau)\ a_\kk\ + h.c.\bigr)
$$
where $k x\equiv k_\eta \eta+\vec k_\perp \vec x_\perp$, 
$\ddk\equiv dk_\eta d^2k_\perp/(2\pi)^3$. 
The mode functions and $\chi$  obey:
\begin{equation}
\ddot f_\kk + 
\bigl({k_\eta^2\over{\tau^2}}+\vec k_\perp^2 + \chi(\tau) +
{1\over{4\tau^2}}\bigr) 
f_\kk=0.\label{eq:mode}
\end{equation}
\begin{equation}
\chi(\tau) = \lambda\Bigl(-v^2+\Phi_i^2(\tau)+{1\over \tau}N \int \ddk
|f_\kk(\tau)|^2\  (1+2\ n_\kk) \Bigr) \label{eq:chi}.
\end{equation}

We notice that when $\chi$ goes negative, the low momentum modes with
$${k_\eta^2 +1/4 \over{\tau^2}}+\vec k_\perp^2 < | \chi |$$ grow
exponentially. However these modes then feed back into the $\chi$
equation and this exponential growth then gets damped. It is these
growing modes that lead to the possiblity of growing domains of DCC's
as well as a modification of the low momentum distribution of
particles from a thermal one. The parameters of the model are fixed by
physical data.  The PCAC condition is $$
\partial_{\mu} A_{\mu}^i (x) \equiv f_{\pi} m_{\pi}^2 \pi^i(x) = H
\pi^i(x). 
$$
In the vacuum state $ \chi_0 \sigma_0= m_{\pi}^2 \sigma_0=H$,
so that $\sigma_0=f_{\pi}$= 92.5 MeV.  The vacuum  gap equation is
$$
m_{\pi}^2 = - \lambda v^2 + \lambda f_{\pi}^2 +
 \lambda N \int_0^{\Lambda} \ddk {1\over 2\sqrt{k^2+m_{\pi}^2}}.
$$
This leads to the mass renormalized gap equation:
\begin{eqnarray}
&&\chi(\tau)-m_{\pi}^2 = -\lambda f_{\pi}^2+ \lambda \Phi_i^2(\tau)
\nonumber \\ 
&&+{\lambda \over \tau}N
\int \ddk
\{|f_\kk(\tau)|^2\  (1+2\ n_\kk) - {1\over 2\sqrt{k^2+m^2}} \}.
\end{eqnarray}
$\lambda$ is chosen to fit low energy scattering data as discussed in
\cite{emil3}.
 
If we assume that the inititially (at $\tau_0 = 1$) the system is in
local thermal equilibrium in a comoving frame we have  
\begin{equation}
n_k = {1 \over e^{\beta_0  E^0 _k} -1}
\end{equation}
where $ \beta_0 = 1/T_0$ and $E^0_k=
\sqrt{{k_\eta^2\over{\tau_0^2}}+\vec k_\perp^2 + \chi(\tau_0)}$.

The initial value of $\chi$ is determined by the equilibrium gap
equation for an initial temperature of $ 200 MeV$ and is $.7 fm^{-2}$
and the initial value of $\sigma$ is just ${H \over \chi_0 }$. The
phase transition in this model occurs at a critical temperature of
$160 MeV$. To get into the unstable domain, we then introduce
fluctuations in the time derivative of the classical field.  We varied
the value of the initial proper time derivative of the sigma field
expectation value and found that for $\tau_0 = 1 fm$ there is a narrow
range of initial values that lead to the growth of instabilities,
namely $.25 < \vert \dot{\sigma} \vert < 1.3$.

\epsfxsize=7cm
\epsfysize=6cm
\centerline{\epsfbox{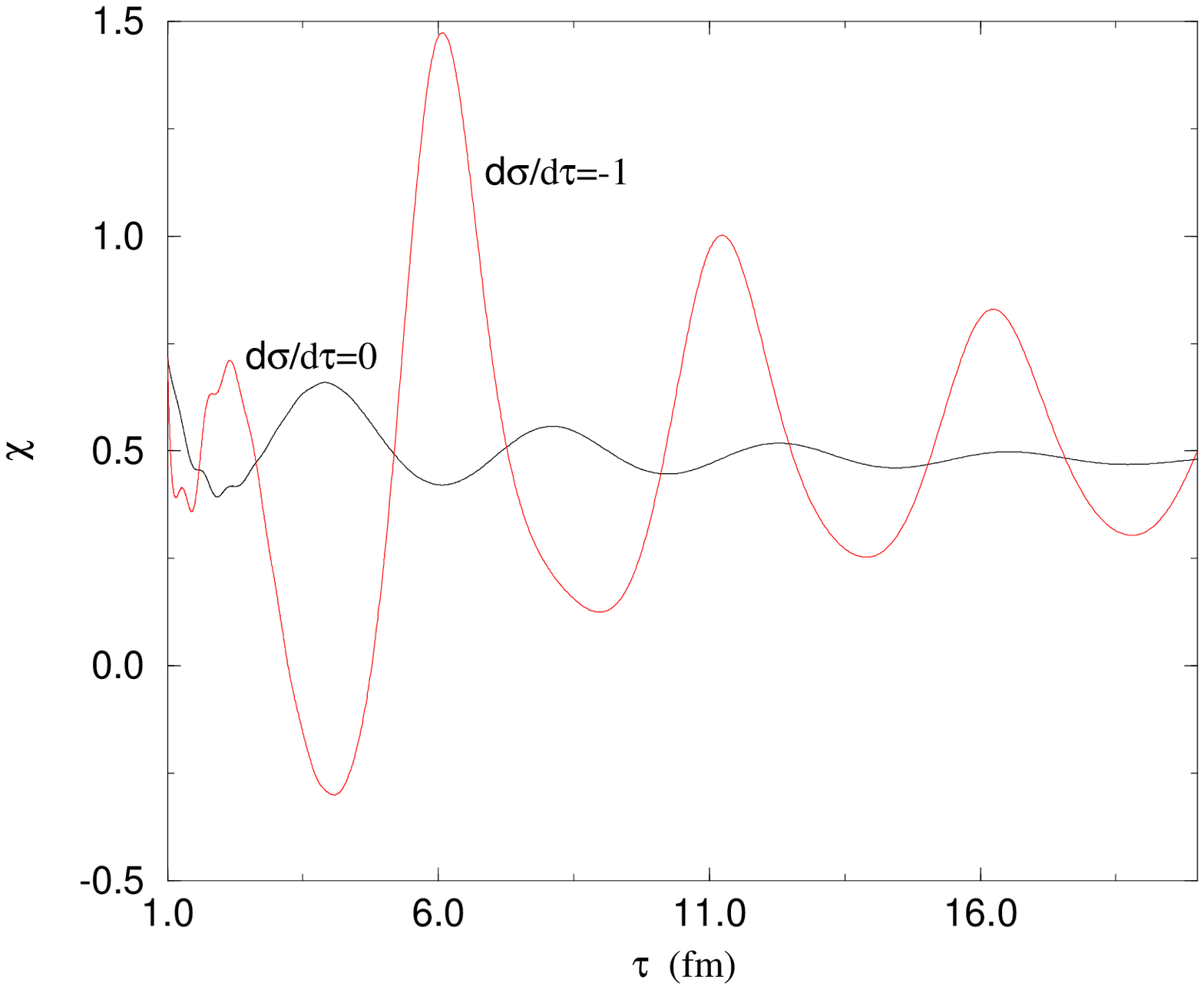}}
\vspace{.35cm}
{FIG. 10.{\small{Proper time evolution of the $\chi$ field for two
different initial values of $\dot{\sigma}$.}}}\\

Fig.10  displays the results of the numerical simulation for the
evolution of $\chi$ (\ref{eq:mode})--(\ref{eq:chi}).
We display the auxiliary field $\chi$ in units of $fm^{-2}$ 
, the classical fields $\Phi$ in units of $fm^{-1}$ 
and the proper time
in units of $fm$   ($ 1fm^{-1} = 197
MeV$) for two simulations, one with an instability
($\dot\sigma|_{\tau_0}=-1$)
and one without  ($\dot\sigma|_{\tau_0}=0$).           

We notice that for both initial conditions, the system eventually
settles down to the broken symmetry vacuum result as a result of the
expansion.  The evolution of the quantities $\sigma$ and $\pi_1$ are
displayed for various initial conditions in fig.11.\\

\vspace{1cm}
\epsfxsize=8.5cm
\epsfysize=9cm
\centerline{\epsfbox{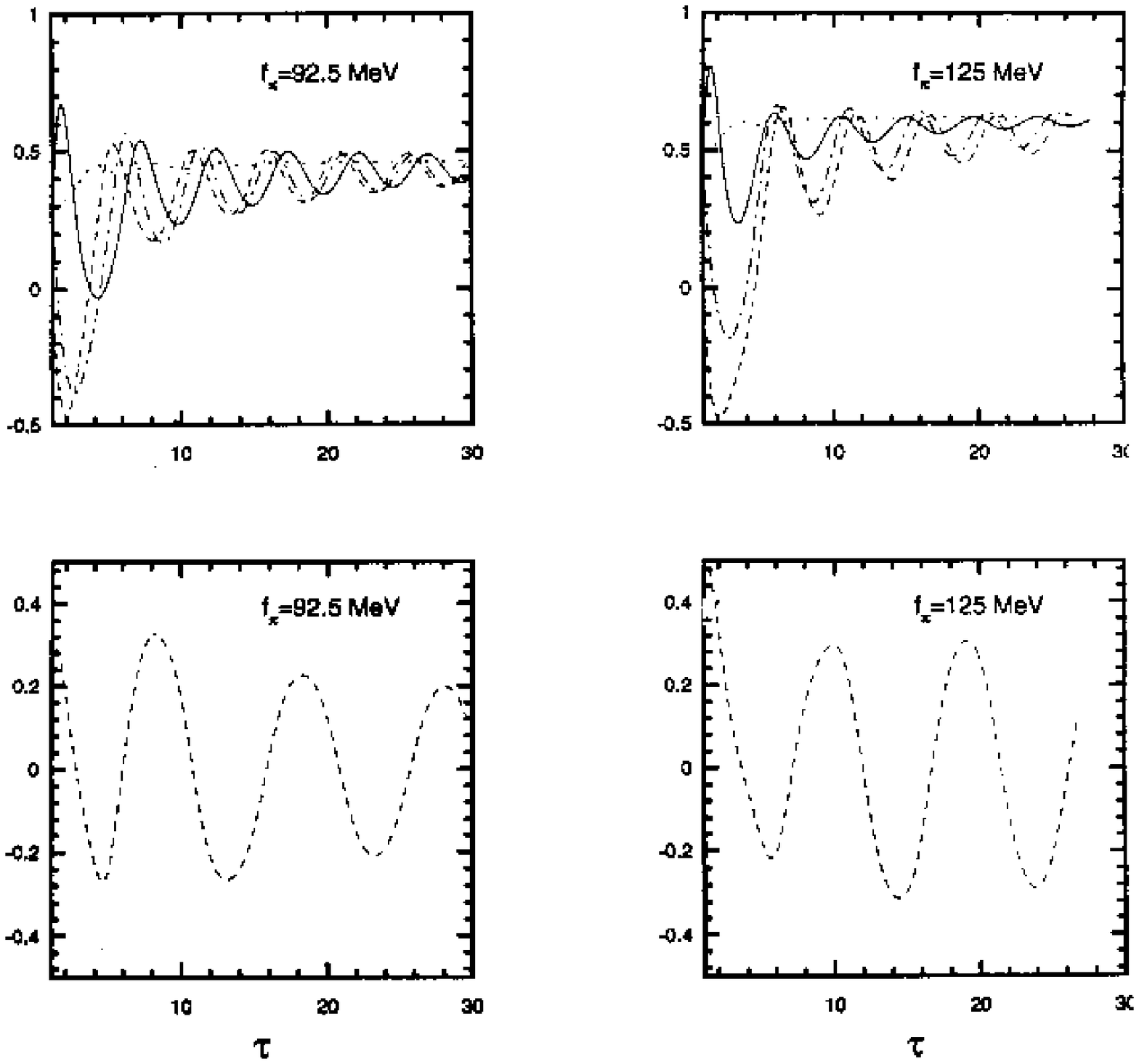}}
\vspace{.35cm}
{FIG. 11{\small{Proper time evolution of the $\sigma$ field  and $\pi$ field
for four different initial conditions with $f_{\pi}=92.5 MeV, 125
MeV$.}}}\\           

To determine the spectrum of particles we introduce the interpolating
number density which is defined by expanding the fields in terms of
mode functions $f^0_k$ which are first order in an adiabatic expansion
of the mode equation.
\begin{equation}
f^0_k = {e^{-iy_k(\tau)} \over \sqrt{2 \omega_k}}; ~~ dy_k/dt =
\omega_k,
\end{equation}
where $\omega_\kk(\tau)\equiv({k_\eta^2/{\tau^2}}+
\vec k_\perp^2 + \chi(\tau))^{1/2}$.
This leads to the alternative expansion of the fields:
\begin{equation}
\phi(\eta,x_\perp,\tau)\equiv{1\over{\tau^{1/2}}} \int
\ddk\bigl(\exp(ikx)
f^0_\kk(\tau)\ a_\kk(\tau)\ + h.c.\bigr)
\end{equation}

The two sets of creation and annihilation operators are connected by 
a Bogoliubov transformation:
\begin{equation}
a_k(\tau) = \alpha(k,\tau) a_k + \beta(k,\tau) a^{\dag}_{-k}.
\end{equation}
$\alpha$ and $\beta$ can be determined from the exact time evolving mode
functions via:
\begin{eqnarray}
\alpha(k,\tau) &= & i (f_k^{0\ast} { \partial f_k \over \partial \tau} -
{\partial f_k^{0 \ast}  \over \partial \tau}f_k) \nonumber \\
\beta(k,\tau) &=& i (f_k^{0} { \partial f_k \over \partial \tau} -
{\partial f_k^{0 }  \over \partial \tau}f_k).
\end{eqnarray}
In terms of the initial distribution of particles $n_0(k)$ and $\beta$
we have:
\begin{eqnarray}
n_k(\tau) &\equiv& f(k_{\eta},k_{\perp},\tau) 
= < a^{\dag}_k (\tau) a_k (\tau) >\nonumber\\
&=& n_0(k) + |\beta(k,\tau)|^2 ( 1+2n_0(k)).
\end{eqnarray}
$n_k(\tau)$ is the adiabatic invariant interpolating phase space
number density which becomes the actual particle number density when
interactions have ceased. When this happens the distribution of
particles is
\begin{equation}
f(k_{\eta}, k_{\perp},\tau) = {d^6 N \over \pi^2 dx_{\perp}^2
dk_{\perp}^2
d\eta dk_{\eta}}.
\label{interp}
\end{equation} 
We now need to relate this quantity to the physical spectra of
particles measured in the lab. At late $\tau$ our system relaxes to
the vacuum and $\chi$ becomes the square of the physical pion mass
$m^2$. As before, we introduce the outgoing pion particle rapidity $y$
and $m_{\perp}= \sqrt{k_{\perp}^2 + m^2}$ defined by the particle
4-momentum in the center of mass coordinate system.  The boost that
takes one from the center of mass coordinates to the comoving frame
where the energy momentum tensor is diagonal is given by $\tanh \eta=
v = z/t$, so that one can define the ``fluid" 4-velocity in the center
of mass frame as
\begin{equation}
u^{\mu} = (\cosh \eta,0,0, \sinh \eta)
\end{equation}
The variable 
\begin{equation}
\omega_k = \sqrt{m_{\perp}^2+{ k_{\eta}^2 \over \tau^2} } \equiv
k^{\mu}u_{\mu}
\end{equation}
 has the meaning of the energy of the particle in the comoving
frame.  
As discussed earlier the 
variables $\tau,\eta$
have as their canonical momenta
\begin{eqnarray}
k_{\tau}= Et/ \tau -k_z z/ \tau \quad \, k_{\eta}  = -Ez + t k_z.
\end{eqnarray}
Changing  variables from $(\eta, k_{\eta})$ to
$(z, y)$ at a fixed $\tau$
we have
\begin{eqnarray}
E{d^3 N \over d^3 k} &=&
{d^3N \over  \pi dy\,dk_{\perp}^2 } = 
\int\pi dz~ dx_{\perp}^2 ~ J ~ f(k_{\eta},
k_{\perp},\tau) \nonumber \\
&& =  A_{\perp} \int dk_{\eta} 
f(k_{\eta}, k_{\perp},\tau)
\end{eqnarray}
This is again converted into the Cooper-Frye form by noting that a
constant $\tau$ surface is
parametrized as: \begin{equation}
d\sigma^{\mu} = A_{\perp} (dz,0,0,dt) 
= A_{\perp}d\eta (\cosh \eta ,0,0,\sinh \eta )
\end{equation}
Thus
\begin{equation}
k^{\mu} d\sigma_{\mu} =  A_{\perp} m_{\perp} \tau \cosh(\eta-y)=
A_{\perp} |
dk_{\eta} |
\end{equation}
and
\begin{equation}
{d^3N \over  \pi dy\,dk_{\perp}^2} = A_{\perp} \int dk_{\eta} 
f(k_{\eta}, k_{\perp},\tau)= \int f(k_{\eta}, k_{\perp},\tau) k^{\mu}
d\sigma_{\mu}
\end{equation}
This reconfirms the idea that the interpolating phase space number
 density plays the role of a classical transport phase space density
 function, as was found in our calculation of pair production from
 strong electric fields \cite{emil2}.

We wish to compare our nonequilibrium calculation with the results of
the hydrodynamical model in the same boost-invariant approximation.
In the hydrodynamical model of heavy ion collisions \cite{fred2}, the
final spectra of pions is given by a combination of the fluid flow and
a local thermal equilibrium distribution in the comoving frame.
\begin{equation}
E {d^3N \over d^3k} = {d^3N \over \pi dk_{\perp}^2 dy} =
\int g(x,k) k^{\mu} d \sigma_{\mu}
\end{equation}
Here $g(x,k)$ is the single particle relativistic phase space
distribution
function. 
When there is local thermal equilibrium of pions at a comoving
temperature $T_c(\tau)$ one has
\begin{equation}
g(x,k) = { g_{\pi} } \{ {\rm exp} [k^{\mu}u_{\mu}/T_c] -1
\}
^{-1}. 
\end{equation}

In Figures 12 and 13 we compare the boost invariant hydrodynamical
result for the transverse momentum distribution using critical
temperatures of $ T_c = 140, 200$ MeV to the two nonequilibrium cases
represented in figure 10.  Figure 12 pertains to the initial condition
${\dot \sigma} |_{\tau_0}=-1$ In this case there is a regime where the
effective mass becomes negative and we see a noticable enhancement of
the low transverse momentum spectra. We have normalized both results
to give the same total center of mass energy $E_{cm}$.  Figure 13
corresponds to the initial condition ${\dot \sigma} |_{\tau_0}=0$.
Here we notice that there is a little enhancement at low transverse
momenta.

\epsfxsize=7cm
\epsfysize=6cm
\centerline{\epsfbox{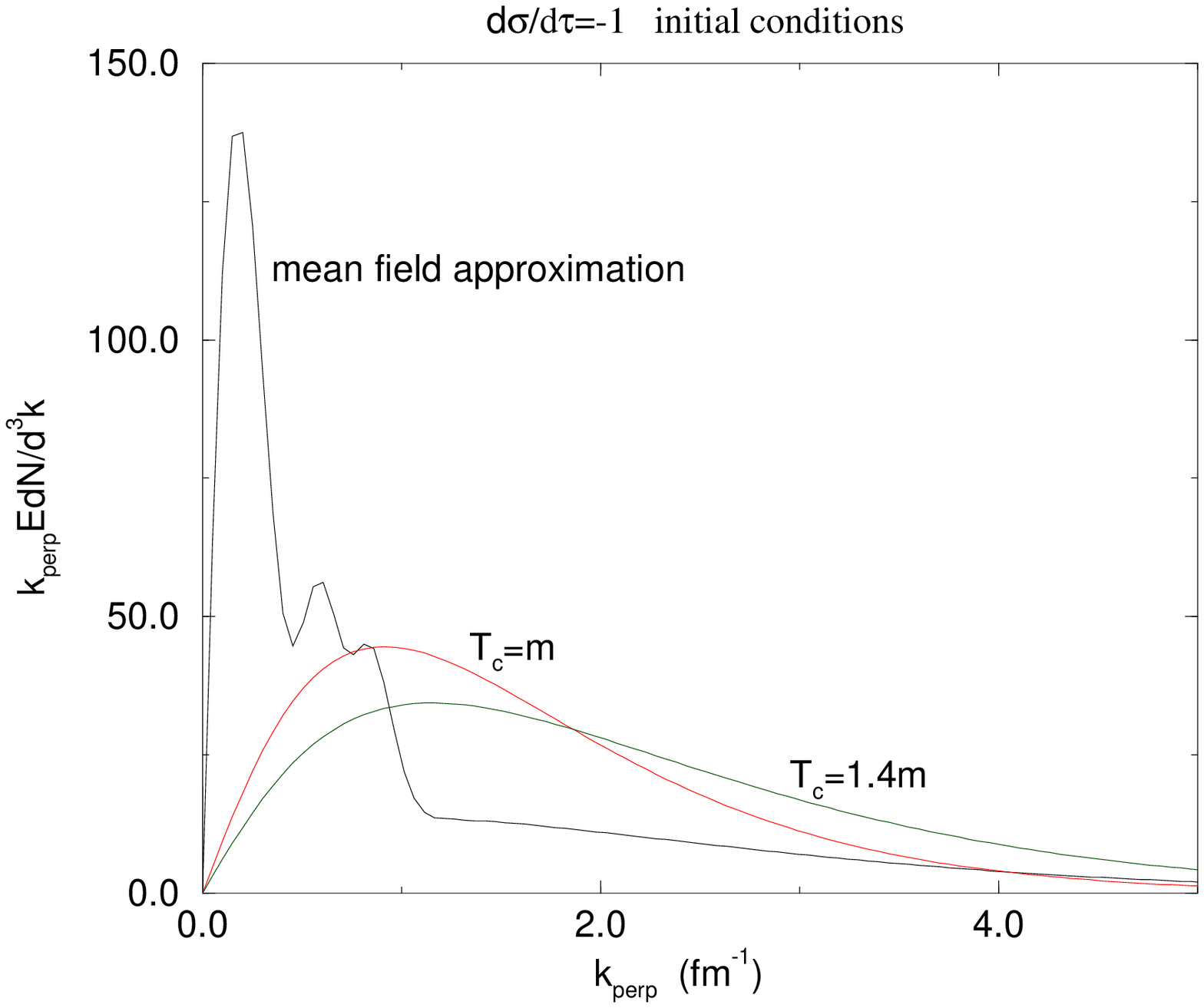}}
\vspace{.35cm}
{FIG. 12. {\small{Single particle transverse momentum distribution for
$\dot{\sigma} =-1$ initial  conditions compared to a local equilibrium
Hydrodynamical calculation with boost invariance.}}}\\ 
\\         
\epsfxsize=7cm
\epsfysize=6cm
\centerline{\epsfbox{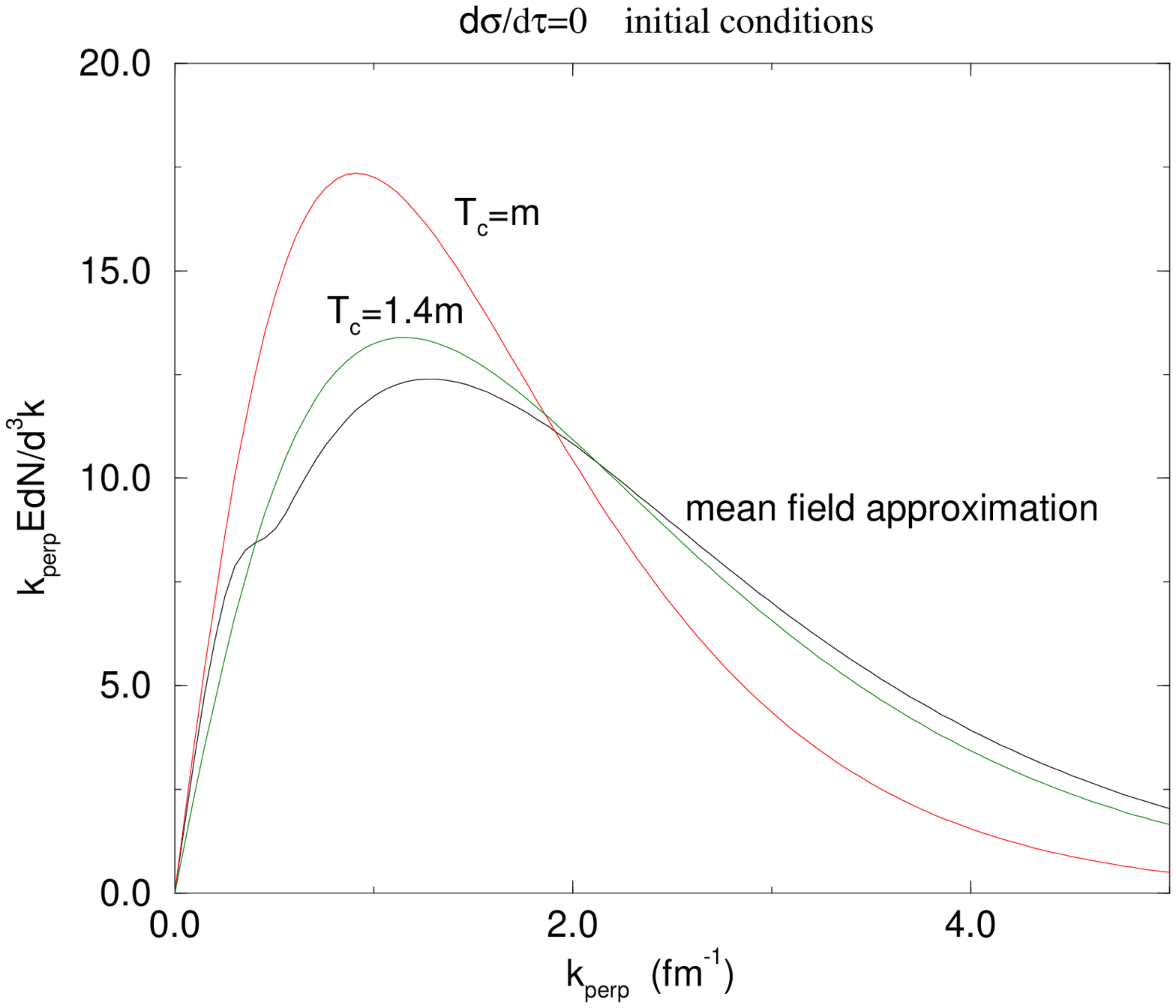}}
\vspace{0cm}
{FIG. 13. {\small{Single particle transverse momentum distribution for
$\dot{\sigma}=0$ initial conditions compared to a local
equilibrium Hydrodynamical calculation with boost invariance.}}}\\          

So we see that a non-equilibrium phase transition taking place during
a time evolving quark-gluon or hadronic plasma can lead to an
enhancement of the low momentum transverse momentum distribution. In
particular, if a Centauro type event is not accompanied by such an
enhancement one would be suspicious of ascribing this event to the
production of disoriented chiral condensates as a result of a rapid
quench.  For the case when we have massless goldstone pions in the
$\sigma$ model ($H=0$) \cite{noneq},then $\chi$ goes to zero at large
times. In that case we find the amusing fact that the equation of
state becomes $p= \varepsilon/3$ at late times. This is true even
though the final particle spectrum is far from thermal
equilibrium. This equation of state is shown in Fig 14. \\

\epsfxsize=7cm
\epsfysize=6cm
\centerline{\epsfbox{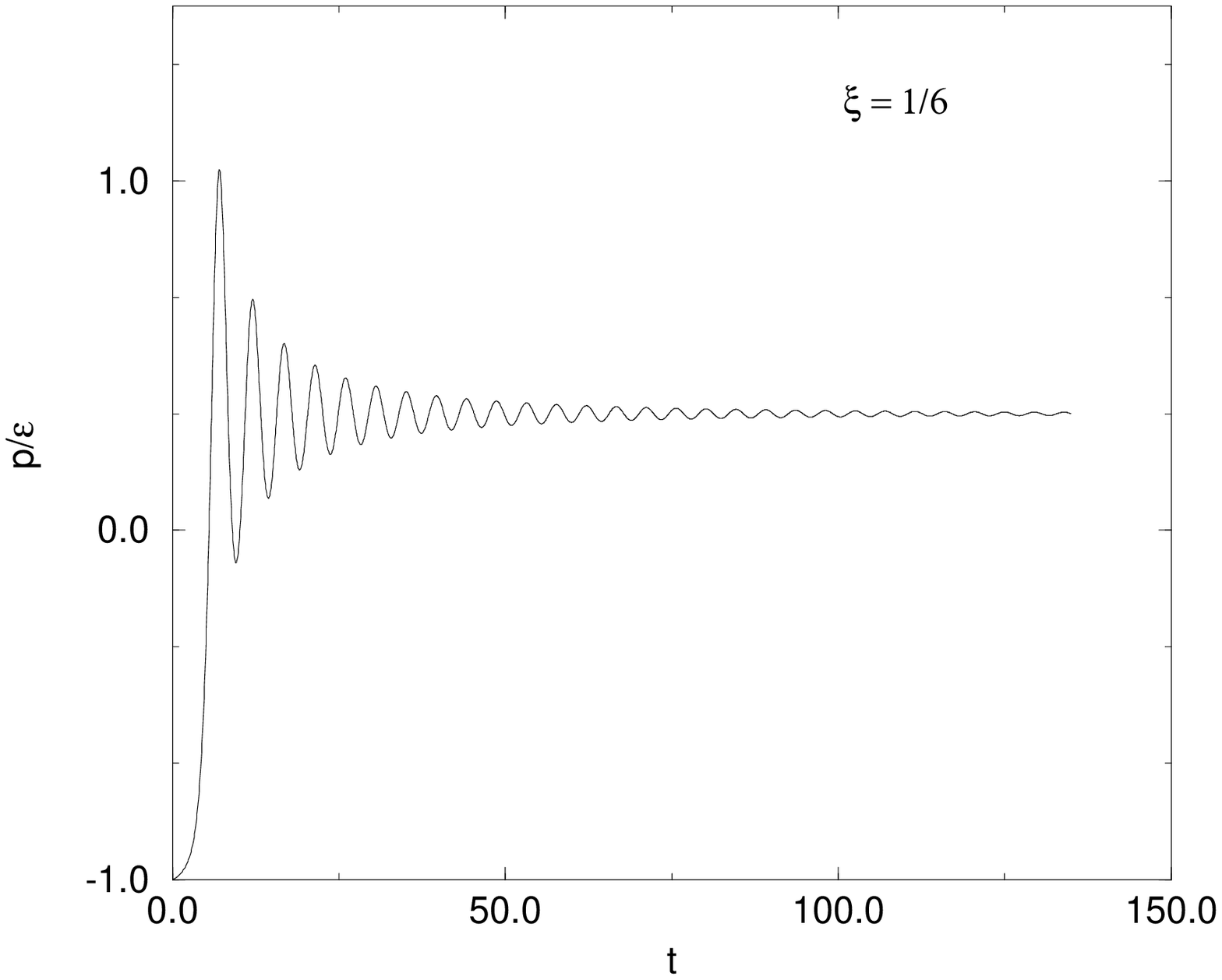}}
\vspace{.2cm}
{FIG. 14. {\small{Equation of state ${p \over \varepsilon}$ as a function
of $\tau$ for the massless $\sigma$ model where we start from a quench.
}}}\\          

\section{ACKNOWLEDGEMENTS}  

The work presented here was done in collaboration with Emil Mottola,
Salman Habib, Yuval Kluger, Juan Pablo Paz, Ben Svetitsky, Judah
Eisenberg, Paul Anderson, So-Young Pi, John Dawson, David Sharp,
Mitchell Feigenbaum, Edmond Schonberg and Graham Frye.  This work was
supported by the Department of Energy.

\vspace{1.0in}

\end{document}

%% file: title2.tex
\catcode`\@=11

\def\maketitle2{\par 
\begingroup
\let\cite\@bylinecite
\def\thefootnote{\fnsymbol{footnote}}%
\twocolumn[\@maketitle2\vskip2pc]%
\thispagestyle{plain}\@thanks
\endgroup
\def\thefootnote{\arabic{footnote}}%
\setcounter{footnote}{0}%
\let\maketitle2\relax \let\@maketitle2\relax
\let\@thanks\relax \let\@authoraddress\relax \let\@title\relax
\let\@date\relax \let\thanks\relax \let\@abstract\relax 
\let\@pacs\relax}

\def\abstract#1{\gdef\@abstract{{\par 
\bgroup
\ifdim\prevdepth=-1000pt \prevdepth0pt\fi
\hsize\columnwidth
\dimen0=-\prevdepth \advance\dimen0 by17.5pt \nointerlineskip
\small\vrule width 0pt height\dimen0 \relax}{~~}#1\egroup}}

\def\pacs#1{\gdef\@pacs{{\par 
\bgroup
\hsize\columnwidth \parindent0pt
\ifdim\prevdepth=-1000pt \prevdepth0pt\fi
\dimen0=-\prevdepth \advance\dimen0 by20pt\nointerlineskip
\egroup} PACS numbers:~#1}}

\def\@maketitle2{
\@preprint
\@title
\ifdim\prevdepth=-1000pt \prevdepth0pt\fi
\@authoraddress
\@date
\begin{list}{}{\leftmargin=0.10753\textwidth \rightmargin=\leftmargin
\itemsep=1pc\partopsep=-1pc}
\item\@abstract
\item\@pacs
\end{list}
}

\catcode`\@=12